\documentclass{article}

\PassOptionsToPackage{numbers, sort, comma, compress}{natbib}

\usepackage[preprint]{neurips_2025}

\usepackage{xcolor}         
\newcommand{\method}{~SOPHY}

\definecolor{codegreen}{rgb}{0,0.6,0}
\definecolor{codegray}{rgb}{0.5,0.5,0.5}
\definecolor{codepurple}{rgb}{0.58,0,0.82}
\definecolor{backcolour}{rgb}{0.97,0.97,0.97}
\usepackage{listings}
\lstdefinestyle{pythonstyle}{
    language=Python,
    backgroundcolor=\color{backcolour},   
    commentstyle=\color{codegreen},
    keywordstyle=\color{magenta},
    numberstyle=\ttfamily\tiny\color{codegray},
    stringstyle=\color{codepurple},
    basicstyle=\ttfamily\tiny,
    breakatwhitespace=false,         
    breaklines=true,                 
    captionpos=b,                    
    keepspaces=true,                 
    numbers=left,                    
    numbersep=5pt,                  
    showspaces=false,                
    showstringspaces=false,
    showtabs=false,                  
    tabsize=2,
    xleftmargin=0.01\textwidth,
    xrightmargin=0.01\textwidth,
}
\lstdefinestyle{markdownstyle}{
    basicstyle=\ttfamily\scriptsize,
    backgroundcolor=\color{backcolour},   
    xleftmargin=0.01\textwidth,
    xrightmargin=0.01\textwidth,
    breakindent=0\dimen0,
    columns=flexible,
    showspaces=false,
    showstringspaces=false,
    breaklines=true,
    breakatwhitespace=true,
    breakautoindent=true,
}

\usepackage{comment}
\usepackage{enumitem}
\makeatletter
\renewcommand{\paragraph}{%
  \@startsection{paragraph}{4}%
  {\z@}{0.25ex \@plus 1ex \@minus .2ex}{-1em}%
  {\normalfont\normalsize\bfseries}%
}
\makeatother

\usepackage[utf8]{inputenc} 
\usepackage[T1]{fontenc}    
\usepackage{url}            
\usepackage{booktabs}       
\usepackage{amsfonts}       
\usepackage{amsmath}        
\usepackage{hyperref}       
\usepackage{nicefrac}       
\usepackage{microtype}      
\usepackage{graphicx}
\usepackage{mathtools}
\usepackage{multirow}
\usepackage[ruled,vlined,linesnumbered]{algorithm2e}
\usepackage{algorithmic}
\usepackage{setspace}
\usepackage{wrapfig}
\usepackage[capitalize]{cleveref}
\usepackage{enumitem}
\usepackage{caption}

\title{\textcolor{teal}{\method}: 
Learning to Generate \textcolor{teal}{S}imulation-Ready \textcolor{teal}{O}bjects with \textcolor{teal}{PHY}sical Materials
}

\author{%
  Junyi Cao\textsuperscript{1, 2} \qquad Evangelos Kalogerakis\textsuperscript{1, 2} \\
  \textsuperscript{1}Technical University of Crete \\
  \textsuperscript{2}University of Massachusetts Amherst \\
}

\makeatletter
\DeclareRobustCommand\onedot{\futurelet\@let@token\@onedot}
\def\@onedot{\ifx\@let@token.\else.\null\fi\xspace}

\def\eg{\emph{e.g}\onedot} 
\def\ie{\emph{i.e}\onedot} 
 
\def\etc{\emph{etc}\onedot}

\makeatother

\begin{document}











\def\support{\mbox{support}}
\def\diag{\mbox{diag}}
\def\rank{\mbox{rank}}
\def\grad{\mbox{\text{grad}}}
\def\dist{\mbox{dist}}
\def\sgn{\mbox{sgn}}
\def\tr{\mbox{tr}}
\def\card{{\mbox{Card}}}

\def\balpha{\mbox{{\boldmath $\alpha$}}}
\def\bbeta{\mbox{{\boldmath $\beta$}}}
\def\bzeta{\mbox{{\boldmath $\zeta$}}}
\def\bgamma{\mbox{{\boldmath $\gamma$}}}
\def\bdelta{\mbox{{\boldmath $\delta$}}}
\def\bmu{\mbox{{\boldmath $\mu$}}}
\def\bftau{\mbox{{\boldmath $\tau$}}}
\def\beps{\mbox{{\boldmath $\epsilon$}}}
\def\blambda{\mbox{{\boldmath $\lambda$}}}
\def\bLambda{\mbox{{\boldmath $\Lambda$}}}
\def\bnu{\mbox{{\boldmath $\nu$}}}
\def\bomega{\mbox{{\boldmath $\omega$}}}
\def\bfeta{\mbox{{\boldmath $\eta$}}}
\def\bsigma{\mbox{{\boldmath $\sigma$}}}
\def\bzeta{\mbox{{\boldmath $\zeta$}}}
\def\bphi{\mbox{{\boldmath $\phi$}}}
\def\bxi{\mbox{{\boldmath $\xi$}}}
\def\bvphi{\mbox{{\boldmath $\phi$}}}
\def\bdelta{\mbox{{\boldmath $\delta$}}}
\def\bvarpi{\mbox{{\boldmath $\varpi$}}}
\def\bvarsigma{\mbox{{\boldmath $\varsigma$}}}
\def\bXi{\mbox{{\boldmath $\Xi$}}}
\def\bmW{\mbox{{\boldmath $\mW$}}}
\def\bmY{\mbox{{\boldmath $\mY$}}}

\def\bPi{\mbox{{\boldmath $\Pi$}}}

\def\bOmega{\mbox{{\boldmath $\Omega$}}}
\def\bDelta{\mbox{{\boldmath $\Delta$}}}
\def\bPi{\mbox{{\boldmath $\Pi$}}}
\def\bPsi{\mbox{{\boldmath $\Psi$}}}
\def\bSigma{\mbox{{\boldmath $\Sigma$}}}
\def\bUpsilon{\mbox{{\boldmath $\Upsilon$}}}

\def\mA{{\mathcal A}}
\def\mB{{\mathcal B}}
\def\mC{{\mathcal C}}
\def\mD{{\mathcal D}}
\def\mE{{\mathcal E}}
\def\mF{{\mathcal F}}
\def\mG{{\mathcal G}}
\def\mH{{\mathcal H}}
\def\mI{{\mathcal I}}
\def\mJ{{\mathcal J}}
\def\mK{{\mathcal K}}
\def\mL{{\mathcal L}}
\def\mM{{\mathcal M}}
\def\mN{{\mathcal N}}
\def\mO{{\mathcal O}}
\def\mP{{\mathcal P}}
\def\mQ{{\mathcal Q}}
\def\mR{{\mathcal R}}
\def\mS{{\mathcal S}}
\def\mT{{\mathcal T}}
\def\mU{{\mathcal U}}
\def\mV{{\mathcal V}}
\def\mW{{\mathcal W}}
\def\mX{{\mathcal X}}
\def\mY{{\mathcal Y}}
\def\mZ{{\mathcal{Z}}}


\def\bmA{{\mathbfcal A}}
\def\bmB{{\mathbfcal B}}
\def\bmC{{\mathbfcal C}}
\def\bmD{{\mathbfcal D}}
\def\bmE{{\mathbfcal E}}
\def\bmF{{\mathbfcal F}}
\def\bmG{{\mathbfcal G}}
\def\bmH{{\mathbfcal H}}
\def\bmI{{\mathbfcal I}}
\def\bmJ{{\mathbfcal J}}
\def\bmK{{\mathbfcal K}}
\def\bmL{{\mathbfcal L}}
\def\bmM{{\mathbfcal M}}
\def\bmN{{\mathbfcal N}}
\def\bmO{{\mathbfcal O}}
\def\bmP{{\mathbfcal P}}
\def\bmQ{{\mathbfcal Q}}
\def\bmR{{\mathbfcal R}}
\def\bmS{{\mathbfcal S}}
\def\bmT{{\mathbfcal T}}
\def\bmU{{\mathbfcal U}}
\def\bmV{{\mathbfcal V}}
\def\bmW{{\mathbfcal W}}
\def\bmX{{\mathbfcal X}}
\def\bmY{{\mathbfcal Y}}
\def\bmZ{{\mathbfcal Z}}

\def\0{{\bf 0}}
\def\1{{\bf 1}}

\def\bA{\boldsymbol{A}}
\def\bB{\boldsymbol{B}}
\def\bC{\boldsymbol{C}}
\def\bD{\boldsymbol{D}}
\def\bE{\boldsymbol{E}}
\def\bF{\boldsymbol{F}}
\def\bG{\boldsymbol{G}}
\def\bH{\boldsymbol{H}}
\def\bI{\boldsymbol{I}}
\def\bJ{\boldsymbol{J}}
\def\bK{\boldsymbol{K}}
\def\bL{\boldsymbol{L}}
\def\bM{\boldsymbol{M}}
\def\bN{\boldsymbol{N}}
\def\bO{\boldsymbol{O}}
\def\bP{\boldsymbol{P}}
\def\bQ{\boldsymbol{Q}}
\def\bR{\boldsymbol{R}}
\def\bS{\boldsymbol{S}}
\def\bT{\boldsymbol{T}}
\def\bU{\boldsymbol{U}}
\def\bV{\boldsymbol{V}}
\def\bW{\boldsymbol{W}}
\def\bX{\boldsymbol{X}}
\def\bY{\boldsymbol{Y}}
\def\bZ{\boldsymbol{Z}}

\def\ba{\boldsymbol{a}}
\def\bb{\boldsymbol{b}}
\def\bc{\boldsymbol{c}}
\def\bd{\boldsymbol{d}}
\def\be{\boldsymbol{e}}
\def\bff{\boldsymbol{f}}
\def\bg{\boldsymbol{g}}
\def\bh{\boldsymbol{h}}
\def\bi{\boldsymbol{i}}
\def\bj{\boldsymbol{j}}
\def\bk{\boldsymbol{k}}
\def\bl{\boldsymbol{l}}
\def\bm{\boldsymbol{m}}
\def\bn{\boldsymbol{n}}
\def\bo{\boldsymbol{o}}
\def\bp{\boldsymbol{p}}
\def\bq{\boldsymbol{q}}
\def\br{\boldsymbol{r}}
\def\bs{\boldsymbol{s}}
\def\bt{\boldsymbol{t}}
\def\bu{\boldsymbol{u}}
\def\bv{\boldsymbol{v}}
\def\bw{\boldsymbol{w}}
\def\bx{\boldsymbol{x}}
\def\by{\boldsymbol{y}}
\def\bz{\boldsymbol{z}}

\def\hy{\hat{y}}
\def\hby{\hat{{\bf y}}}

\def\mmE{{\mathbb E}}
\def\mmP{{\mathrm P}}
\def\mmB{{\mathrm B}}
\def\mmR{{\mathbb R}}
\def\mmV{{\mathbb V}}
\def\mmN{{\mathbb N}}
\def\mmZ{{\mathbb Z}}
\def\mMLr{{\mM_{\leq k}}}

\def\tC{\tilde{C}}
\def\tk{\tilde{r}}
\def\tJ{\tilde{J}}
\def\tbx{\tilde{\bx}}
\def\tbK{\tilde{\bK}}
\def\tL{\tilde{L}}
\def\tbPi{\mbox{{\boldmath $\tilde{\Pi}$}}}
\def\tw{{\bf \tilde{w}}}

\def\barx{\bar{\bx}}

\def\pd{{\succ\0}}
\def\psd{{\succeq\0}}
\def\vphi{\varphi}
\def\trsp{{\sf T}}

\def\mRMD{{\mathrm{D}}}
\def \DKL{{D_{KL}}}
\def\st{{\mathrm{s.t.}}}
\def\nth{{\mathrm{th}}}

\def\st{{\mathrm{s.t.}}}
\def\tr{\mathrm{tr}}
\def\grad{{\mathrm{grad}}}

\newtheorem{coll}{Corollary}
\newtheorem{deftn}{Definition}
\newtheorem{thm}{Theorem}
\newtheorem{prop}{Proposition}
\newtheorem{ass}{Assumption}




\maketitle
\vspace{-4mm}
\includegraphics[width=\linewidth]{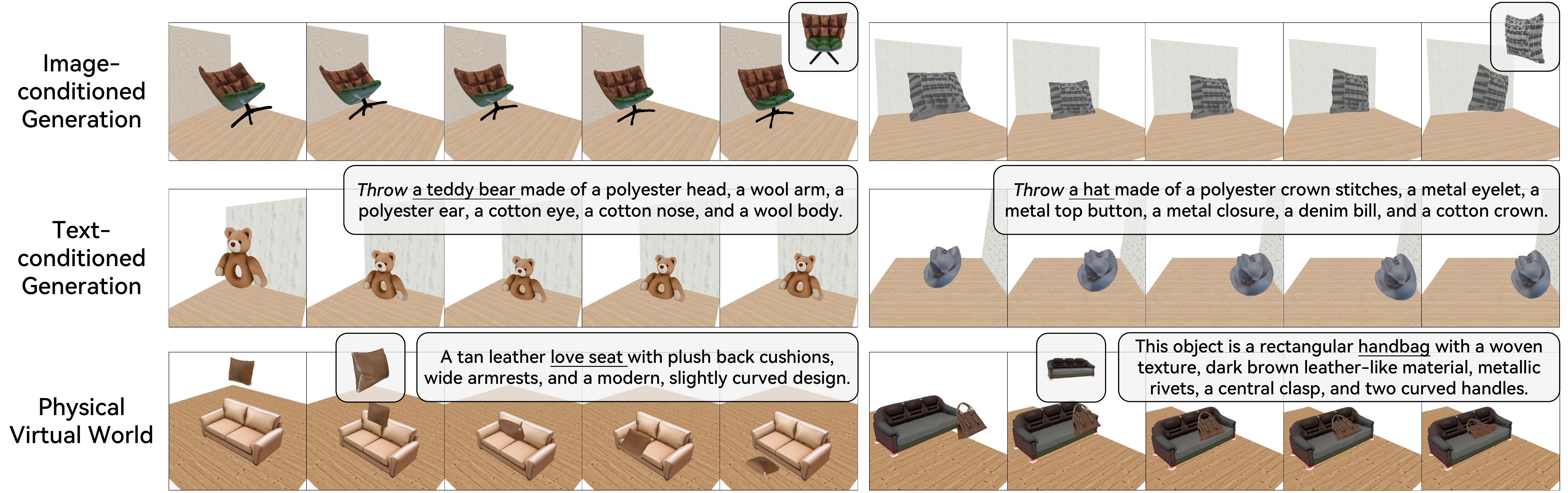}
\captionof{figure}{\method~is a physics-aware generative model designed for creating 3D objects that possess plausible geometries, textures, and physical properties, making them ready for simulation. 
(\emph{Top})
Our method
can generate these assets from a single image;  
(\emph{middle}) or from
a text prompt (conditions are shown as insets in the top right corner of each generated object).
(\emph{Bottom}) We further showcase\method's potential for constructing a physically accurate virtual world by combining various image and text conditions to generate diverse objects.}
\label{fig:teaser}

\begin{abstract}
We present\method, a generative model for 3D physics-aware shape synthesis. Unlike existing 3D generative models that focus solely on static geometry or 4D models that produce physics-agnostic animations, our method jointly synthesizes shape, texture, and material properties related to physics-grounded dynamics, making the generated objects ready for simulations and interactive, dynamic environments. 
To train our model, we introduce a dataset of 3D objects annotated with detailed physical material attributes, along with an efficient pipeline for material annotation. Our method enables applications such as text-driven generation of interactive, physics-aware 3D objects and single-image reconstruction of physically plausible shapes. Furthermore, our experiments show that jointly modeling shape and material properties enhances the realism and fidelity of the generated shapes, improving performance on both generative geometry and physical plausibility. Our project page is available at: \url{https://xjay18.github.io/SOPHY_page}.
\end{abstract}
\section{Introduction}
\label{sec:intro}

Generating high-quality 3D assets that can be incorporated in interactive virtual worlds or manufactured into functional 3D objects is a fundamental challenge in generative AI and digital content creation. Unfortunately, current generative models of 3D shapes~\cite{chen20243dtopia,zhao2025hunyuan3d,siddiqui2024meta} are limited to generating only static geometry and textures. Although there is significant effort in building 4D generative models for both geometry and motion synthesis~\cite{wu2024cat4d,sun2024dimensionx,liangdiffusion4d}, current methods are limited to fixed animations and are incapable of producing motions resulting from complex, dynamic object interactions. 

In this work, we introduce a diffusion-based generative model for physics-aware 3D object synthesis, capable of generating 3D assets with detailed shape, texture, and, most importantly, physical material attributes governing kinematic object deformations (\eg, elastic deformations, plastic softening) and frictional interactions. Our approach produces physics-aware, interactive 3D objects, which are helpful for downstream applications such as physically based simulation, robotic interaction, and manufacturing. There are various challenges to developing such a generative approach. First, current 3D and 4D datasets do not contain physical material attributes for objects, making it hard to train, or even fine-tune such physics-aware generative models. Second, it is unclear how to jointly model the interplay of shape and material in 3D asset generation. 
Clearly, material attributes are not independent of geometric shapes, as repeatedly observed in prior works of physical material recognition from shapes ~\cite{li20223d,lin2018learning,slim20233dcompat,ahmed20253dcompat200}. For example, a chair’s backrest made of thin rods or strips is usually metal, while a thicker, block-like backrest is often made of fabric or soft padding. 

Our work aims to address the above challenges. First, we introduce a semi-automatic annotation pipeline, combining Vision Language Model (VLM)'s guidance with iterative expert (mechanical engineer) feedback for more efficient object annotation with material properties. Second, we introduce a generative model that builds upon recent advancements in latent diffusion models, yet incorporates novel insights on representing 3D objects as compact latent codes capturing shape geometry, albedo color, and material properties related to elasticity and plasticity, as well as capturing their interdependence through cross-attention blocks. Additionally, we integrate a texture enhancement module as a post-processing step that refines the generated albedo color, resulting in a high-quality texture map for improved appearance. Overall, we observe that modeling shape and material properties together enhances the realism and fidelity of the geometry in the generated shapes. 
Our contributions are:
\begin{itemize}[noitemsep,topsep=0pt,leftmargin=*]
    \item We present a semi-automatic pipeline to obtain annotations describing object physical behavior, and a new dataset of 3K objects and 15K parts annotated with detailed physical material properties.
    \item We introduce an autoencoder for representing physics-aware 3D objects into compact latents, along with a diffusion-based model that jointly synthesizes geometry, texture, and physical materials.
    \item We showcase applications of our framework for synthesizing simulation-ready objects based on texts or images, as well as  physics-aware virtual worlds filled with objects generated by our method.
\end{itemize}

\section{Related Work}
\label{sec:related_work}
\paragraph{3D asset generation.} 
Over the recent years, we have witnessed an explosion of 3D generative models capable of generating detailed geometry and texture. The advances have been driven by expressive neural representations, such as NeRFs ~\cite{mildenhall2020nerf,Poole:2022:Dreamfusion}, Gaussian splats \cite{kerbl20233d}, signed distance fields \cite{chen2018implicit_decoder,Park_2019_CVPR}, occupancy fields \cite{Mescheder19},
and numerous other representations for capturing shape or/and texture \cite{oechsle2019texture,shen2021deep,hui2022wavelet,Lin:2023:Magic3D, xu2024grm,chen2024meshanything,chen2024meshanythingv2,jun2023shapegeneratingconditional3d,zhang20233dshape2vecset,petrov2024gem3d,georgiou2025im2surftex,hong2024lrm,xu2024instantmesh,bala2024edify} to name a few. We also refer readers to the recent surveys of 3D generative model~\cite{li2024advances3dgenerationsurvey,wang2024diffusionmodels3dvision}.
Unfortunately, the vast majority of existing 3D generative models neglect the physical material properties of the generated 3D objects, limiting their applicability to interactive and simulation-based environments. To address this gap, recent research has begun incorporating physical priors and constraints into the generative pipeline. 
For example, DiffuseBot~\cite{wang2023diffusebot} evaluates generated 3D robot models based on their simulation performance and refines the sampling distribution to favor more successful designs. Atlas3D~\cite{chen2024atlas3d} and Phys-Comp~\cite{guo2024physically} enforce static equilibrium constraints during shape optimization, ensuring that generated objects remain stable under gravity.
Despite these advances, all these prior works assume uniform or limited material properties, restricting their ability to generate diverse and physically intricate 3D assets. In contrast, our approach jointly optimizes 3D shape, albedo color, and material properties with a network that captures their natural dependencies.

\paragraph{Material-annotated 3D datasets.}
The advances in 3D generative models have also been led by the development of influential 3D datasets such as ShapeNet~\cite{chang2015shapenet}
and Objaverse~\cite{deitke2023objaverse}. 
More recent datasets, including ABO~\cite{collins2022abo}, Matsynth~\cite{vecchio2024matsynth}, and BlenderVault~\cite{litman2025materialfusion}, incorporate surface texture information to train models for 3D generation and more plausible, photorealistic rendering.
Despite this progress, a critical gap remains in the availability of detailed physical material information essential for accurately modeling material properties for physics-based simulations. 
While datasets such as ShapeNet-Mat~\cite{lin2018learning}, 3DCoMPaT~\cite{li20223d} and its extensions~\cite{slim20233dcompat,ahmed20253dcompat200} attempt to address this limitation by providing part-level material labels, their annotations are coarse and unsuitable for direct use in physical simulators.
To bridge this gap, we introduce a new dataset consisting of 3K objects spanning 12 shape categories, each annotated with precise part-level physical material properties.

\paragraph{4D content generation.}
4D generation aims to create dynamic 3D content that aligns with input conditions such as images, text, or videos. Existing approaches primarily follow a data-driven pipeline~\cite{lee2024vividdream,watson2024controlling,wu2024cat4d,liangdiffusion4d,li20244k4dgen,sun2025egd}, leveraging off-the-shelf image or video diffusion models to generate dynamic scenes. For example, DreamGaussian4D~\cite{ren2023dreamgaussian4d} adopts Stable Video Diffusion~\cite{blattmann2023stable} to animate 3D Gaussian splats (3DGS)~\cite{kerbl20233d} reconstructed from a single image. STAG4D~\cite{zeng2024stag4d} initializes multi-view images using image-to-image diffusion~\cite{shi2023zero123++} anchored on input video frames from a text-to-video module. However, these methods rely on pre-trained generative models that lack physical understanding, often resulting in physically unrealistic motions~\cite{bansal2024videophy}. 
Another line of research~\cite{gao2025partrm,liu2025riganything,sun2025armo} focuses on part-level motion based on articulation, yet is able to accommodate only rigid motion of parts.
To address more general physics-aware dynamics, 
PhysGaussian~\cite{xie2023physgaussian} and PhysDreamer~\cite{PhysDreamer} integrate the Material Point Method (MPM)~\cite{jiang2016material} to animate 3DGS representations. 
Building on these efforts, several recent studies~\cite{lin2024phys4dgen,tan2024physmotion,chen2025physgen3d,liu2025unleashing} have focused on physical dynamics generation from a single image. They typically follow a two-stage pipeline to obtain simulation-ready objects: generating 3D shapes first and then assigning them with homogeneous physical attributes by manual selection or estimation from VLMs.
Despite offering an intuitive framework, they overlook the interdependence between geometries and material properties as they split shape generation and material estimation into two steps. Consequently, the assigned attributes may not accurately reflect the material heterogeneity and the tightly coupled nature of geometry observed in the real world.
In contrast, our method introduces a learning-based approach for simulation-ready object generation. By jointly modeling shape geometries and material properties in an end-to-end framework, our method ensures greater physical realism and geometric coherence.

\section{Material Annotation}
\label{sec:material_annotation}
We first describe the procedure for creating a dataset of 3D objects annotated with detailed physical material properties.
Specifically, we annotate shapes with the material parameters used in the Material Point Method (MPM)~\cite{sulsky1994particle,jiang2016material,hu2018moving}, a popular simulator known for its effectiveness in handling complex simulations
of behaviors and inertia effects inherent in solids ~\cite{xie2023physgaussian,pacnerf,cao2024neuma,PhysDreamer}, including but not limited to elastic and plastic deformation of solids and frictional interactions. The material parameters include: (a) Young's modulus, (b) Poisson’s ratio, (c) yield stress, (d) friction angle, (e) density, and (f) material behavior type (also known as material model), including pure elastic deformation~\cite{wineman2005some}, plastic deformation with softening (damage)~\cite{wang2020material}, plastic deformation without softening~\cite{wang2019simulation}, and granular deformation~\cite{klar2016drucker}.

Unfortunately, no datasets of 3D objects exist with such detailed material properties available. The most related dataset is 3DCoMPaT~\cite{li20223d} along with its subsequent versions (3DCoMPaT++~\cite{slim20233dcompat}, 3DCoMPaT200~\cite{ahmed20253dcompat200}), which provide a rather coarse labeling of materials for 3D shape parts, such as ``plastic'', ``fabric'', ``metal'', \etc. While these categories provide some insight into the general physical characteristics of objects, they are often too broad for accurate simulation in physics engines. For example, a ``plastic'' part made out of rigid PVC is much stiffer (\ie, has a much higher Young's modulus) than a flexible plastic material (\eg, LDPE). Unfortunately, such properties can be provided only by domain experts or manufacturers, and are difficult to interpret or estimate for annotators without a strong physics background and experience on commonly used material types in objects. Hiring experts to label these properties for every single component of every 3D object in a large database would be extremely cumbersome. Thus, we devised a semi-automatic pipeline that combines material property annotation by VLMs, followed by iterative expert feedback and verification. 

\begin{wrapfigure}[10]{r}{0.64\textwidth}
\vspace{-4mm}
\centering
\begin{minipage}{0.63\textwidth}
    \centering    \includegraphics[width=1\linewidth]{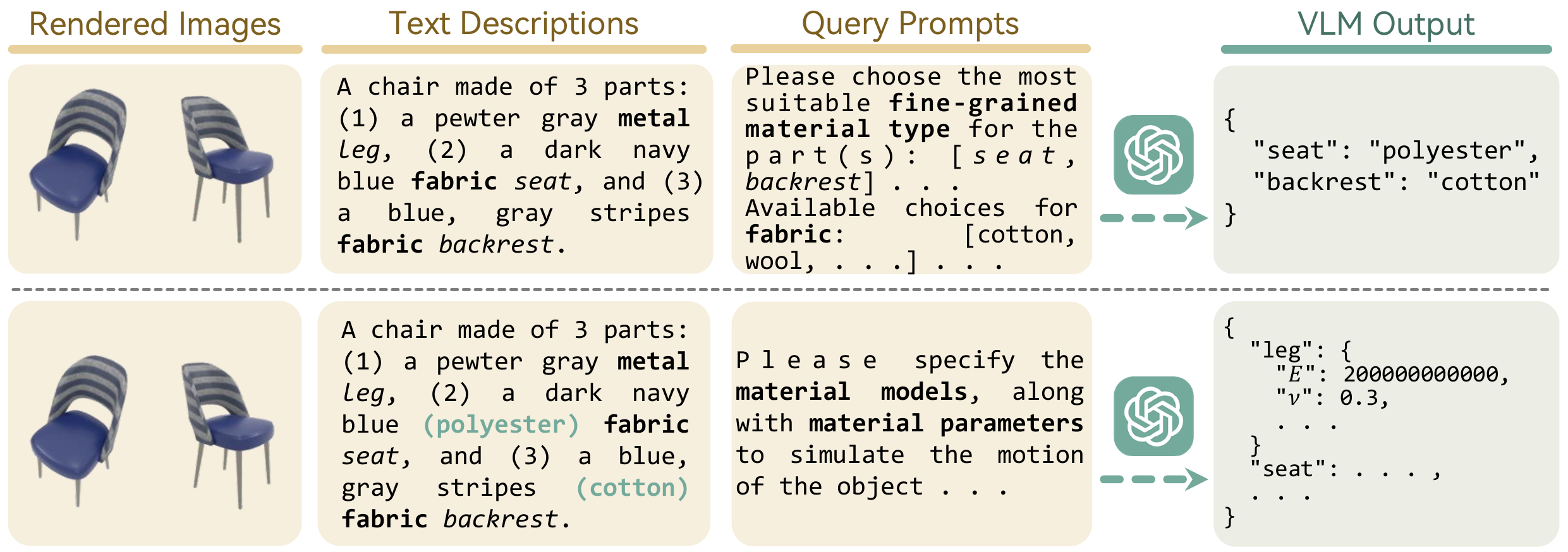}
    \caption{\bf VLM-guided material annotation of shape parts.}    
    \label{fig:physical_reasoning}
\end{minipage}
\end{wrapfigure}

\subsection{VLM-guided Material Property Proposals}
\label{sec:vlm_guidance}
VLMs have recently become prominent in physical reasoning~\cite{zhen20243dvla,chen2024commonsense,lai2024vision,zhao2024automated,lin2024phys4dgen,liu2024physgen} due to their extensive knowledge bases built on multi-modal data. Thus, we leveraged their zero-shot reasoning ability to give an initial estimation of material properties for input 3D shapes. We started by choosing $12$ shape categories from 3DCoMPaT200~\cite{ahmed20253dcompat200}, which contained a variety of material compositions and behaviors capable of elastic or plastic deformation, such as bags, pillows, and chairs. We skipped categories where objects behave rigidly (\eg, cabinets, faucets).
As shown in \Cref{fig:physical_reasoning}(top), we provided a popular VLM (ChatGPT-4o~\cite{hurst2024gpt4o}) with (a) two automatically generated, rendered views of each textured object in the selected categories, (b) a textual description of the shape, including its object category, part tag, coarse material label, and color, as provided by 3DCoMPaT200, (c) a list of available fine-grained material categories \eg, for ``plastic'', we include sub-types such as polypropylene, rigid or flexible PVC, and so on. We provide this fine-grained list in~\Cref{sec:supp_fine_grained_types}. Our text prompt asks the VLM to identify the most plausible fine-grained material category per part. Then for each part, in a second round, we further prompted the VLM to provide the most likely material models and parameters (\eg, Young's modulus, Poisson’s ratio, yield stress, and friction angle) based on a similar textual description, including this time the fine-grained material category as shown in~\Cref{fig:physical_reasoning}(bottom). This two-round prompt provided better material property prediction based on our expert verification. 

\begin{wrapfigure}[9]{r}{0.57\textwidth}
\vspace{-10mm}
\centering
\begin{minipage}{0.56\textwidth}
\centering    
\includegraphics[width=1\linewidth]{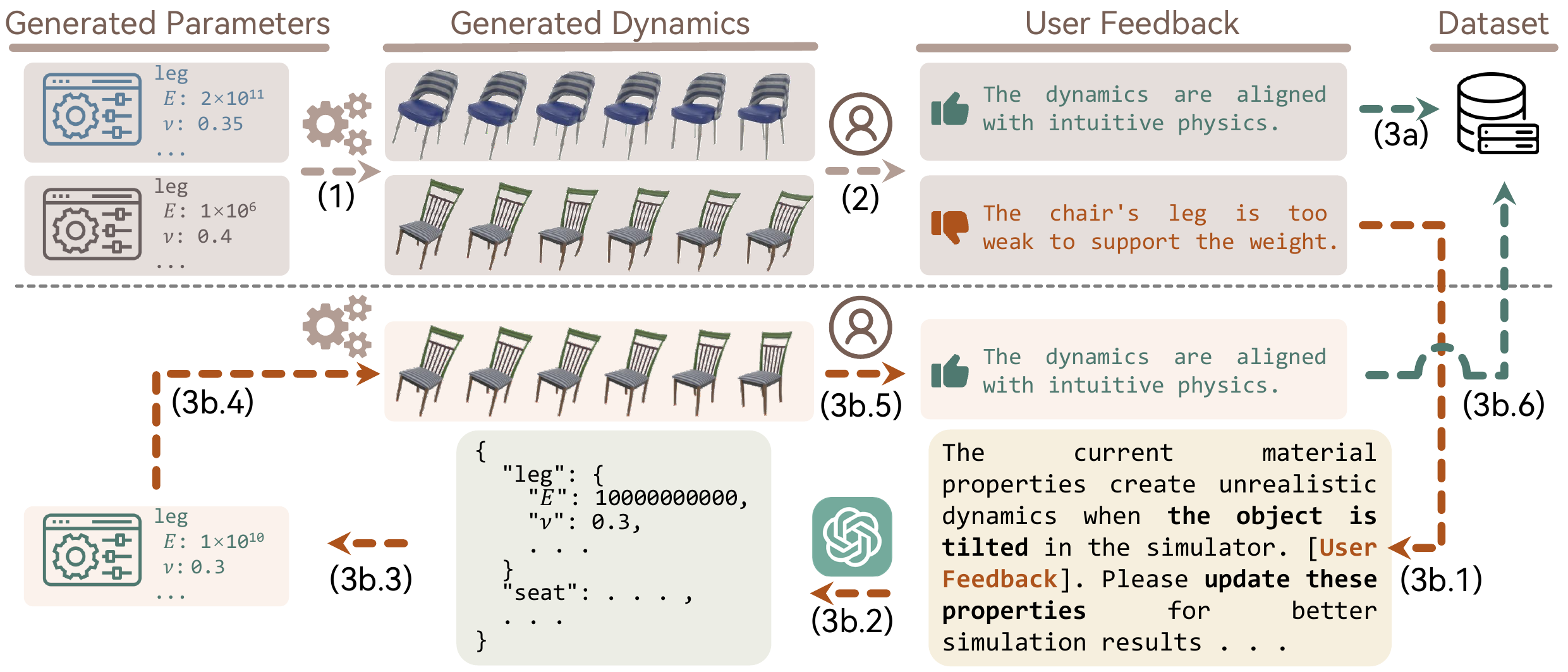}
\vspace{-4mm}
\caption{\bf Expert verification of material annotations.}
\label{fig:sim_based_verification}
\end{minipage}
\end{wrapfigure}

\subsection{Expert Verification and Feedback}
\label{sec:verification}
Although VLMs often offer reasonable initial estimates of material properties, they are not always reliable. To address this issue, we develop a pipeline that generates simulation videos of objects based on the material properties provided by the VLM, then ask experts with mechanical engineering backgrounds to assess their physical plausibility. 

\paragraph{Test scenarios.} We created five test scenarios to simulate object dynamics: (1) \emph{Drop}ping an object from a certain height; (2) \emph{Throw}ing an object in a certain direction; (3) \emph{Tilt}ing an object; (4) \emph{Drag}ging an object; (5) Applying a short-term, time-variant (\eg, \emph{wind}) force. We obtain object motions under these scenarios using a particle-based simulator~\cite{zong2023neural} that we extended to model heterogeneous materials per particle, since our dataset often contains objects whose components are made of different materials. Then, we render object sequences with a high-performance renderer~\cite{Laine2020diffrast,kerbl20233d} and obtain the simulation videos.
After collecting these videos, we enlisted five mechanical engineering graduates to evaluate the physical plausibility of each video and instructed them to reach a consensus on their judgment. As shown in~\Cref{fig:sim_based_verification}(2)-(3a), for simulated objects deemed as realistic by this group, we store their material properties in our dataset. 
For objects deemed to have unrealistic motion, we ask the group to provide feedback to the VLM, specifying which part of the object seems to move in an implausible manner and why. 
This allows us to re-query VLMs for material parameters, as demonstrated in  \Cref{fig:sim_based_verification}(3b.1)-(3b.5), and generate new simulation videos automatically based on the updated parameters. This process is iterative until satisfactory simulation results are achieved.

\subsection{Dataset Summary}
\label{sec:dataset}
After the above verification stage, we obtained $3,004$ 3D models in $12$ object categories originating from the 3DCoMPaT200~\cite{ahmed20253dcompat200} dataset. Every shape part is labeled with physical material properties and detailed material categories, resulting in a total of $15,575$ labeled parts. We split our dataset into training, validation, and test sets, following the 3DCoMPaT200 splits -- the total number of samples in each split is $2462$, $180$, and $362$, respectively.
We provide more information about our dataset, including data statistics and sample simulation sequences in~\Cref{sec:supp_dataset_details}.

\section{Generative Model}

Our proposed generative model,\method, is based on Stable Diffusion~\cite{rombach2022high}, which generates data by denoising a compressed feature space (latent space). To extract this latent space, we process our shapes along with the color and physical material parameters through a variational autoencoder (Section \ref{sec:autoencoding}). 
Then, a latent diffusion model is trained to jointly model the distribution of object geometries, colors, and material properties in this latent space (Section \ref{sec:diffusion}). The rationale behind jointly modeling these properties together is that they are strongly interdependent on each other \ie, certain shapes are strongly correlated with the use of specific materials and colors \eg, thin cantilever-shaped chair bases are associated with the use of metal and have a metallic gray color appearance.
Modeling these properties independently of each other would yield unlikely or impossible materials for sampled geometries. After training our diffusion model, it can be sampled to generate new latent codes, which are then decoded to obtain shapes, compatible colors and physical material properties.

\begin{figure*}[t]
    \centering
    \includegraphics[width=\linewidth]{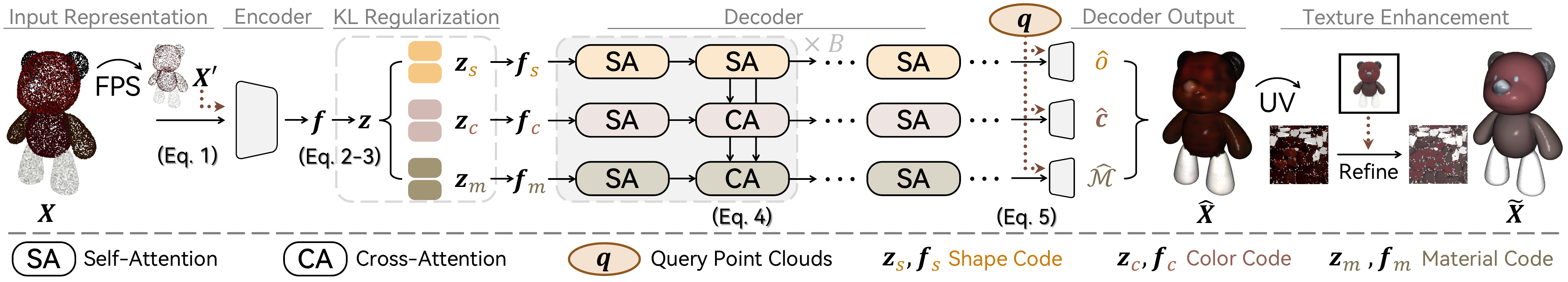}
    \caption{\bf Our pipeline for compressing objects to compact shape, texture, and material codes.}
    \label{fig:generative_model}
    \vspace{-6mm}
\end{figure*}

\subsection{Shape, Texture and Material Autoencoder}
\label{sec:autoencoding}
Our autoencoder
(\Cref{fig:generative_model})
compresses input 3D objects into compact latent codes that encode shape, color (albedo texture) and physical material properties. We discuss it below in more detail.

\paragraph{Input representation.}
We represent an object as a dense surface point cloud 
$\bP = \{\bp_j\}_{j=1}^{N}$, 
where $\bp_j$ is a 3D point position ($N=2048$ in our implementation). The input to our autoencoder is a color- and material-augmented
point cloud
$\bX = \{\bx_j\}_{j=1}^{N}$, where each entry
$\bx_j$ contains the following per-point information: 
(a) 3D position $\bp_j$, (b) RGB color $\bc_j$, (c) a material property vector $\bm_j$
with the following $9$ entries:
(i) logarithm of Young's modulus $E_j$ (a scalar), (ii) Poisson’s ratio $\nu_j$ (a scalar), (iii) logarithm of yield stress $\sigma_j$ (a scalar), (iv) friction angle $\phi_j$ (a scalar), (v) density $\rho_j$ (a scalar), and (vi) a $4$-dimensional learnable embedding $\bmu_j$ of material behavior type.

Note that we use a logarithmic scale for Young's modulus and yield stress due to their extremely wide value ranges.  For example, Young's modulus may vary from $10^{-3}$ GPa for very soft materials to $10^{3}$ GPa for very stiff materials.
We represent material information per point rather than per part to accommodate the general case where a semantic part is not made of a homogeneous material.

\paragraph{Encoder.} 
To aggregate the per-point information from the augmented point cloud, we design a set-to-set network inspired by 3DShape2VecSet~\cite{zhang20233dshape2vecset} as our encoder.
We first sub-sample a smaller point cloud $\bP'$ with $M=512$ fewer points from the original point cloud through furthest point sampling (FPS)~\cite{pointnet++}, and augment it with color and material information to obtain $\bX'= \{\bx'_i\}_{i=1}^{M}$.
Then, we use cross attention to produce the object's latent representation:
\begin{equation}
    \{\bff_i\}_{i=1}^M = \mathrm{CrossAttn}
    \left(
    g\big(\{\bx'_i\}_{i=1}^M\big)
    ,
    g\big(\{\bx_j\}_{j=1}^N\big)
    \right), 
\label{eq:encoder}
\end{equation}
where $g$ is a linear layer projecting the augmented point cloud into the embedding space $\mmR^C$ ($C=512$ in our implementation) and $\{\bff_i\}_{i=1}^M$ is a set of latent codes.

\paragraph{KL regularization.} 
The above representation is quite high-dimensional. We thus seek to compress it towards a more compact latent code. Following Stable Diffusion~\cite{rombach2022high} and 3DShape2VecSet~\cite{zhang20233dshape2vecset}, we adopt a variational autoencoder (VAE) regularized with the KL-divergence to achieve this effect.
We first use two fully connected (FC) layers to project each latent code $\bff_i$ to mean and variance:
\begin{equation}
\bff_i^\mu = \mathrm{FC}_\mu(\bff_i),\ \bff_i^\sigma = \mathrm{FC}_\sigma(\bff_i),
\end{equation}
where $\bff_i^\mu,\bff_i^\sigma\in\mmR^{C_0}$ and $C_0\ll C$ ($C_0=24$ in our implementation). Then, we use the reparameterization sampling and obtain a smaller latent code $\bz_i\in\mmR^{C_0}$:
\begin{equation}
\label{eq:reparameterization}
\bz_i = \bff_i^\mu + \epsilon \cdot \bff_i^\sigma ,\ \epsilon \sim \mN(0,1),
\end{equation}
which enables us to train diffusion models on a lower-dimensional space later. Finally, we project $\bz_i$ back into the original embedding space $\mmR^{C}$ with another FC layer.

\paragraph{Decoder.} One possible decoder design follows the approach of 3DShape2VecSet~\cite{zhang20233dshape2vecset}, where latent codes are first processed through a series of self-attention blocks, then for a query point $\bq \in \mathbb{R}^{3}$, its occupancy is determined by interpolating the latent codes based on the query position and transforming the result through a fully connected layer. This design could be naturally extended to also decode color and material information per query point. However, we observed worse performance using this approach, as color and material information are not meaningful for non-occupied query points (\ie, points outside the shape volume).

Empirically, we achieved better results with an alternative design. We first decode geometry in terms of occupancy values, then decode color, and finally decode material properties, but only for query points classified as occupied. Performance was further improved when we explicitly split the latent codes $\bz_i$ into three sub-codes: the shape code $\bz_{i,s}$ for occupancy decoding, the color code $\bz_{i,c}$ for texture decoding, and the material code $\bz_{i,m}$ for material property decoding. Each sub-code has a shape of $M \times \frac{C_0}{3}$. To decode these components, we introduce three dedicated branches, each responsible for processing one of the sub-codes (see~\Cref{fig:generative_model}).
We also apply cross-attention layers to the color and material branches, enabling them to attend to information from previously decoded latents. This design models the natural dependency of color on geometry and material properties on both geometry and color, which is inspired by the workflow commonly used in 3D asset creation, where artists typically start by defining the object's shape, then apply textures, and finally assign physical attributes.
Specifically, the cross-attention in our decoder is formulated as follows:
\begin{equation}
\begin{gathered}
    \{ \bff^{(l)}_{i,s} \} = \mathrm{SelfAttn}\big(
    \{ \bff^{(l-1)}_{i,s} \}
    \big),\ 
    \{ \bff^{(l)}_{i,c} \} = \mathrm{CrossAttn}\big(
    \{ \bff^{(l-1)}_{i,c} \}, 
    \{ \bff^{(l-1)}_{i,s} \}
    \big), \\ 
    \{ \bff^{(l)}_{i,m} \} = \mathrm{CrossAttn}\big(
    \{ \bff^{(l-1)}_{i,m} \}, \{ \big[  \bff^{(l-1)}_{i,s}, \bff^{(l-1)}_{i,c} \big] \}
    \big),
\end{gathered}
\end{equation}
where $l$ is the layer index, and $[\cdot,\cdot]$ denotes concatenation. 

\paragraph{Decoder output.} Given a query point $\bq \in \mathbb{R}^{3}$, the occupancy values are decoded by interpolating the shape sub-codes from the last layer (layer $L$) and processing them through an FC block:
\begin{equation}
O(\bq)\!\!=\!\!
\mathrm{FC} \Big(\!
\sum\limits_i^{M}\! \mathrm{Softmax}\! \big( \frac{q(
\bff_{\bq} ) \cdot k( \bff^{(L)}_{i,s} )} { \sqrt{C}} \big) v(\bff^{(L)}_{i,s})
\!\Big),
\label{eq:interp2}
\end{equation}
where $\bff_{\bq}$ is a feature vector obtained by processing the query point through our encoder described in \Cref{eq:encoder}. The functions $q(\cdot), k(\cdot), v(\cdot)$ are the query-key-value linear transformations used in attention \cite{vaswani2017attention}. By disentangling the latent codes, we ensure that occupancy is determined solely based on the shape sub-codes relevant to this task.

Color and material properties are decoded similarly, each using its own FC block and interpolation over the corresponding color and material sub-codes. For color prediction, we apply normalization after the FC block to ensure the output falls within the $[0,1]$ range. Young's modulus and yield stress are always positive, so the FC blocks predict their values on a logarithmic scale. 
The friction angle has a range of $[0, \pi/2]$, so we apply a sigmoid activation after the FC block and scale the output by $\pi/2$.
The categorical material behavior type is predicted using a softmax activation after the FC block.
We store the material properties only for query points predicted as occupied. For color, we only predict the values for query points sampled on the mesh surface obtained by marching cubes~\cite{lorensen1987marching}. 

\paragraph{Autoencoder training.} 
While training our variational autoencoder, we jointly optimize a combination of loss functions involving 3D occupancies, colors, and material properties. Specifically, we minimize a weighted sum of the following losses:
 (a) occupancy loss $\mL_o$ (binary cross-entropy),
 (b) color loss $\mL_c$ ($\ell_1$ norm),
 (c) Young's modulus loss $\mL_E$ ($\ell_1$ norm applied to logarithmic values),
 (d) Poisson’s ratio loss $\mL_{\nu}$ ($\ell_1$ norm),
 (e) yield stress loss $\mL_{\sigma}$ ($\ell_1$ norm applied to logarithmic values),
 (f) friction angle loss $\mL_{\phi}$ ($\ell_1$ norm),
 (g) density loss $\mL_{\rho}$ ($\ell_1$ norm),
 (h) material model loss $\mL_M$ (cross-entropy),
 (i) a regularization loss $\mL_{r}$ imposing a KL penalty towards a standard normal distribution on the latent codes, as typically used in VAEs~\cite{kingma2013auto}. We note that the color and material losses are computed only for training query points on or inside the object's surface. For off-surface training points, we assign color and material properties by copying them from their nearest surface points. The combined loss function is:
\begin{equation}
\label{eq:loss}
\mL = \mL_o + \sum_{\omega}\lambda_\omega\mL_{\omega},\quad \omega=\{c, E, \nu, \sigma, \phi, \rho, M, r\},
\end{equation}
where $\lambda_\omega$ are weight parameters for balancing different terms.
To further enhance training, we leverage the pretrained 3DShape2VecSet model~\cite{zhang20233dshape2vecset}, which was trained on ShapeNet~\cite{chang2015shapenet} (a larger dataset containing 55K shapes) using occupancy supervision alone. Specifically, we initialize all network layers shared with 3DShape2VecSet, including the cross-attention weights on point positions in our encoder, the self-attention weights on shape sub-codes, and the occupancy decoder weights, using their pretrained values. This provides a small boost compared to training our model from scratch on our smaller 3K-shape dataset. We also note that we adopt the latent set representation of 3DShape2VecSet~\cite{zhang20233dshape2vecset} in our autoencoder due to its simplicity and efficiency in compressing 3D objects. Our main insight is the joint modeling of shape, texture, and physical properties, and this concept is applicable beyond this specific representation.

\paragraph{Texture enhancement.} As shown in~\Cref{fig:generative_model}, the decoded object $\hat{\bX}$ exhibits a rather flat and blurry visual appearance due to insufficient texture resolution inherited by the vertex-based color representation. Thus, we refine its texture based on renderings of the input shape (in the auto-encoding setting) or conditioned signals (in the generative setting). Specifically, we first generate a UV map, unwrapping the generated object, and then bake its per-vertex color information into a 2D texture based on the UV layout. Subsequently, we adopt the texture enhancement method of Paint3D~\cite{zeng2024paint3d} to refine the 2D texture for improved visual quality. This process serves as an optional post-processing step for our autoencoder. In general, it takes around 1 minute to refine the texture of an object. 

\subsection{Diffusion}
\label{sec:diffusion}

Generating a simulation-ready object involves executing the reverse process of a diffusion that progressively transforms Gaussian noise in the latent space of objects into target latent codes:
\mbox{$\bZ = \{\bz_i\}_{i=1}^M$}.
Sampling is performed by solving the stochastic differential equations from the EDM diffusion pipeline~\cite{karras2022elucidating}. Once a latent code is sampled, we pass it through our trained decoder to extract occupancy values on a dense $\mathbb{R}^3$ grid. These values are then converted into a mesh using marching cubes, then color and material properties are decoded at densely sampled mesh points.

\paragraph{Training.} To train the diffusion model, we use the loss:
\begin{equation}
\mL_\mathrm{edm} = \mmE_{\hat{\bZ}\sim p_{data}}\mmE_{\bN\sim\mN(\boldsymbol{0},\sigma_t^2\bI)}\|D(\hat{\bZ}+\bN,\sigma_t, \bc) -\hat{\bZ}\|_2^2,
\end{equation}
where $D$ is the denoising network, $\hat{\bZ}$ are training latent codes, $\bN$ are added Gaussian noises,
$\sigma_t$ denotes the noise level, and $\bc$ is a signal for conditioning. We experimented with two input conditions to our denoiser: (a) an RGB image of a target object, where $\bc$ represents here the extracted features from a pre-trained image encoder (``DINOv2-ViT-B/14''~\cite{oquab2024dinov}), (b) a text prompt, where $\bc$ represents features from a pre-trained text encoder (``CLIP-ViT-L/14''~\cite{radford2021learning}). The denoising network consists of alternating self-attention layers for processing the latent representations
and cross-attention layers for incorporating the signals $\bc$. Details on the denoiser architecture and conditional signals are provided in~\Cref{sec:supp_denoiser} and~\cref{sec:supp_conditional_signals}, respectively.

\section{Experiments}
\label{sec:experiments}

We evaluate\method~on its autoencoder effectiveness (Section \ref{section:autoenc}), its generative capabilities for image-conditioned generation (Section \ref{sec:image-to-4D})
and text-conditioned generation (Section \ref{sec:text-to-4D}) of simulation-ready 3D objects. All our comparisons were performed in the same test split of our dataset, described in Section \ref{sec:dataset}.

\begin{figure}[!t]
\centering
\begin{minipage}{.47\textwidth}
\centering
\small
\makeatletter\def\@captype{table}\makeatother
\caption{{\bf Quantitative comparison in the auto-encoding setting.} The upper, middle, and lower sections display the average metrics for predictions of material, color, and shape, respectively.}
\label{tab:auto_encoding_metrics}
\scalebox{.85}{
\setlength{\tabcolsep}{1mm}{
\begin{tabular}{r|c|c|c|c}
\toprule
Metric & B. Dec. & w/o CA & Fused & SOPHY \\
\midrule
M.B. Acc(\%) $\uparrow$ & 71.04 & 92.77 & 93.23 &\bf 93.55 \\
MAE-$\log(E)$ $\downarrow$ & 1.18 & 0.50& 0.47 & \bf 0.45 \\
MAE-$\nu$($\times 10^{-2}$) $\downarrow$ & 4.10 &  3.06 &\bf 2.98  & 3.06\\
MAE-$\log(\sigma)$ $\downarrow$ &  1.07 & 0.41 & 0.32 & \bf 0.29 \\
MAE-$\phi$ ($\times 10^{-2}$) $\downarrow$ &   5.45 & 1.89 & \bf 1.22 & 1.28 \\
MAE-$\rho$ $\downarrow$ & 0.16 & 0.14 & 0.13 & \bf 0.09 \\
Sim-CD ($\times 10^{-3}$) $\downarrow$ &53.84 & 17.47& 10.05& \bf 8.72\\
\midrule
MAE-$c$($\times 10^{-2}$) $\downarrow$ & 8.75 & 8.47 & 8.35 & \bf 8.13\\
\midrule
IoU(\%) $\uparrow$ & 90.69 & 90.75&  90.66& \bf 90.89\\
CD($\times 10^{-4}$) $\downarrow$ & 3.51 &3.34 & 3.30 & \bf 3.02 \\
F-Score(\%) $\uparrow$ & 93.65 & 93.77 & 93.79 & \bf 93.85 \\
\bottomrule
\end{tabular}
}}
\end{minipage}
\hspace{4mm}
\begin{minipage}{.48\textwidth}
\centering
\small
\makeatletter\def\@captype{table}\makeatother
\caption{\textbf{Image-to-4D generation.} ``B. Dec.'' and ``B. Perc.'' denote the decoder-based and perception-based baselines, respectively.}
\label{tab:image_to_4d_metrics}
\scalebox{.84}{
\setlength{\tabcolsep}{1mm}{
\begin{tabular}{r|c|c|c|c|c}
\toprule
Metric & DG4D & Free4D & B. Dec. & B. Perc. & SOPHY \\
\midrule
VideoPhy $\uparrow$ & -0.28 & -0.21 & -0.15 & 
0.21 & \bf 0.43\\
CLIP $\uparrow$ & 0.71 & 0.74 & 0.73 & 0.74 & \bf 0.77 \\
Time (min) $\downarrow$ & 10 & 73 & \bf 5 & 7 & \bf 5 \\
\bottomrule
\end{tabular}
}}
\vspace{2mm}
\centering
\small
\makeatletter\def\@captype{table}\makeatother
\caption{\textbf{Text-to-4D generation.} ``B. Dec.'' and ``B. Perc.'' denote the decoder-based and perception-based baselines, respectively.}
\label{tab:text_to_4d_metrics}
\scalebox{.84}{
\setlength{\tabcolsep}{0.8mm}{
\begin{tabular}{r|c|c|c|c|c}
\toprule
Metric & STAG4D & Free4D & B. Dec. & B. Perc. & SOPHY \\
\midrule
VideoPhy $\uparrow$ & -1.23 & 0.07 & -0.18 & 0.58 & \bf 0.76 \\
CLIP $\uparrow$& 0.13 & 0.17 & 0.15 & \bf 0.18 & \bf 0.18 \\
Time (min) $\downarrow$ & 66 & 73 & \bf 5 & 7 & \bf 5 \\
\bottomrule
\end{tabular}
}}
\end{minipage}
\end{figure}

\subsection{Auto-encoding Evaluation}
\label{section:autoenc}
In terms of auto-encoding evaluation, our goal is to check how well we are able to recover a test 3D shape, along with its color and material properties, given its input representation. This evaluation is common in 3D latent-based generative models~\cite{zhang20233dshape2vecset}, since it is imperative for the autoencoder to capture latent spaces that can generalize to novel inputs.

\paragraph{Competing methods.} We stress that there is no existing generative model matching our setting of joint shape and physical material synthesis, thus, we compare with variant designs for our autoencoder:

\noindent \textit{(a) baseline}, or in short ``\emph{B. Dec.}'',
is a model that excludes color and material properties from the generation process \ie, it generates a 3D shape, then predicts color conditioned on the shape through a decoder, and then the material through another decoder. The choice of this baseline attempts to answer the question of whether there is any benefit of incorporating the physical materials in the generation process \ie, whether it is simply better to generate the 3D shape first, then guess its most likely texture and material discriminatively. For this baseline, we use 3DShape2VecSet as the generative model, trained on the same split as our method. We use a texture decoder, which decodes its latent shape representation with a set of self-attention blocks to per-point colors. We use one more decoder to decode the latent shape representation to material properties, including a cross-attention block for conditioning on colors. The number of total parameters remains comparable to our model.

\noindent \textit{(b) w/o CA} is a degraded variant of our proposed\method. It discards the cross-attention blocks used in the color and material decoder branch. This variant is equivalent to decoding our latent sub-codes with three non-interacting branches for occupancy, color, and material predictions.

\noindent  \textit{(c) Fused} is another variant of\method~that uses a unified representation for the latent code, without separating it into the shape, color, and material sub-codes. The decoder infers the per-query occupancy, color, and material properties conditioned on this single latent code.

\paragraph{Metrics.}
For material and color prediction, we report the classification accuracy of the material behavior type (M.B. Acc.) and the mean absolute error (MAE) of all our material parameters. 
In addition, we report Chamfer distance (Sim-CD) between densely sampled points of the predicted shapes and ground-truth ones under the deformed states computed by our MPM simulator along the whole simulation trajectory for the dropping test scenario (\Cref{sec:verification}). 
Finally, we report purely geometric measures for the rest state of the reconstructed test shape compared to the ground truth. We use the standard metrics of Chamfer Distance (CD), volumetric Intersection-over-Union (IoU), and F-score, as used in prior work~\cite{zhang20233dshape2vecset}. We average all values for each object in our test set.

\begin{figure*}[t]
\vspace{-2mm}
    \centering
    \includegraphics[width=\linewidth]{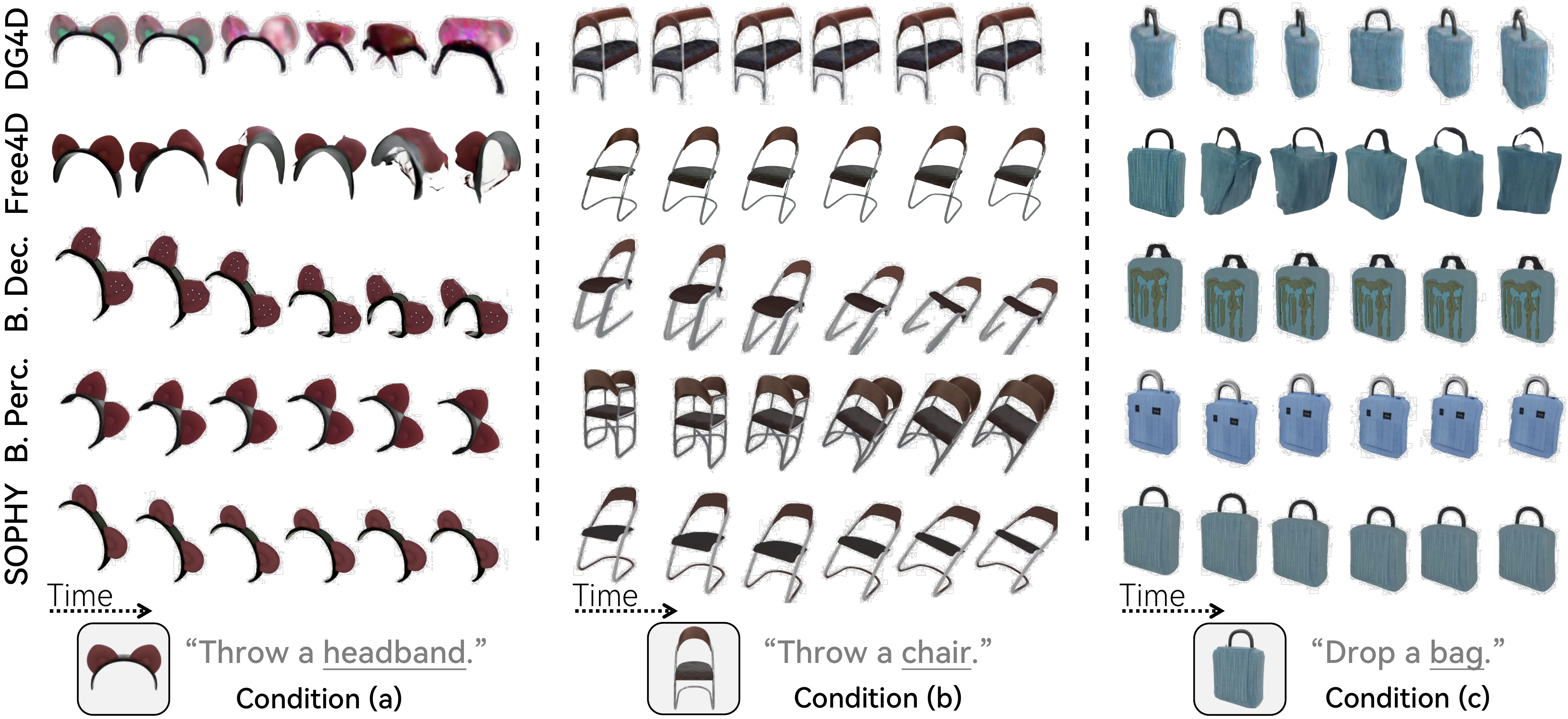}
    \caption{{\bf Qualitative comparison in the image-to-4D setting.} 
    The gray texts next to the conditioned images describe the desired dynamics. Note that the objects fall onto a ground plane, which is not shown here for clarity.}
    \label{fig:image_cond_results}
    \vspace{-6mm}
\end{figure*}

\paragraph{Results.}
\Cref{tab:auto_encoding_metrics} presents quantitative comparisons in the auto-encoding setting. Compared to the baseline model, jointly modeling shape, color, and material in a shared embedding space significantly improves material metrics, with additional, more modest gains in color and geometry evaluation metrics. Notably, the material behavior type classification is improved by more than $20\%$, while Sim-CD is reduced by a factor of $5\times$. This performance gain supports our hypothesis that geometry, color, and material attributes are strongly correlated and should be jointly modeled in the generative process. Compared to not using cross-attention in the decoder (``w/o CA'') model, we see that our full model has $2\times$ lower Chamfer distance during simulation, and still slightly better performance in all other metrics. Finally, we observe that the ``Fused'' variant, 
has a tiny edge over our model in Poisson's ratio and friction angle prediction. Yet, it performs significantly worse in terms of Sim-CD, and worse in all other measures. We suspect this behavior is due to an uneven network capacity allocation across occupancy, color, and material within a single fused latent code. 

\subsection{Image-to-4D}
\label{sec:image-to-4D}
We now discuss applying\method~to generate 4D dynamics given a single input RGB image. Specifically, given a 3D object with physics material properties generated by our method conditioned on the input image, we plug it into a virtual 3D environment with other objects or primitives e.g., ground planes, walls, and so on, and animate it based on its material properties and interactions with the environment. To evaluate in this setting, we render a 2D image from each object of our test split, provide it to our trained diffusion model for generation, then simulate the generated object using the test scenarios of Section \ref{sec:verification} to create 4D scenes.

\paragraph{Competing approaches.}
We first compare\method~with conventional image-to-4D methods, which generate deforming 3DGS over time without any explicit physics-based representation. We examine DreamGaussian4D (DG4D)~\cite{ren2023dreamgaussian4d}, a pioneering work in this field, along with Free4D~\cite{liu2025free4d}, which is one of the state-of-the-art methods that distills pre-trained foundation models for consistent 4D generation.
We also compare with our ``baseline'' discussed in the previous section. In addition, we consider another physics-aware baseline inspired by recent work~\cite{lin2024phys4dgen,chen2025physgen3d}, which adopts a two-stage framework for simulation-ready object generation (see~\Cref{sec:related_work}), and design a training-free baseline, which we call ``\emph{B. Perc}''. Specifically, this alternative baseline uses the perceptual models~\cite{ma2024find,hurst2024gpt4o} for material property estimation given an off-the-shelf 3D generation model~\cite{xiang2024structured}. After obtaining the material properties, we use the same simulator to create dynamics as\method~does. We provide more details about this alternative baseline in~\Cref{sec:supp_perception_based_baseline}.

\paragraph{Metrics.} Given rendered images of the 4D scenes generated by any of the above compared methods, we leverage VideoPhy~\cite{bansal2024videophy}, a state-of-the-art VLM trained based on human annotations, to evaluate whether the generated videos align with real-world physics. The method produces a per-scene score that is uncalibrated for different scenarios. Thus, following~\cite{tan2024physmotion}, we apply $z$-score normalization to calibrate the scores across all methods for each scene and report the average values. A positive $z$-score indicates a method performs better than average, while negative means worse. Additionally, we use the CLIP~\cite{radford2021learning} score to evaluate the consistency between the generated results and the input conditions. We also report the average time to create a video for an object starting from an image.

\paragraph{Results.}
In~\Cref{tab:image_to_4d_metrics},\method~outperforms the other methods, demonstrating the highest z-score for alignment with real-world physics based on VideoPhy. For CLIP scores, our method is better aligned with the conditioned signals. 
We also show qualitative results for a few samples for test scenes in~\Cref{fig:image_cond_results}. 
We observe that DG4D and Free4D struggle to generate physically plausible animations, \eg, the headband size fluctuates over time, as they overlook physical constraints. Both baselines, ``B. Dec.'' and ``B. Perc.'', generate geometries and physical materials independently. Thus, they often create physically implausible objects (\eg, strange chairs in~\Cref{fig:image_cond_results}(b)) or predict inappropriate materials for shapes (\eg, too stiff materials for a bag that looked like it was made of fabric in~\Cref{fig:image_cond_results}(c)). In contrast,\method~produces more realistic dynamics, aligning closely with the given images.

\begin{figure*}[t]
    \centering
    \includegraphics[width=\linewidth]{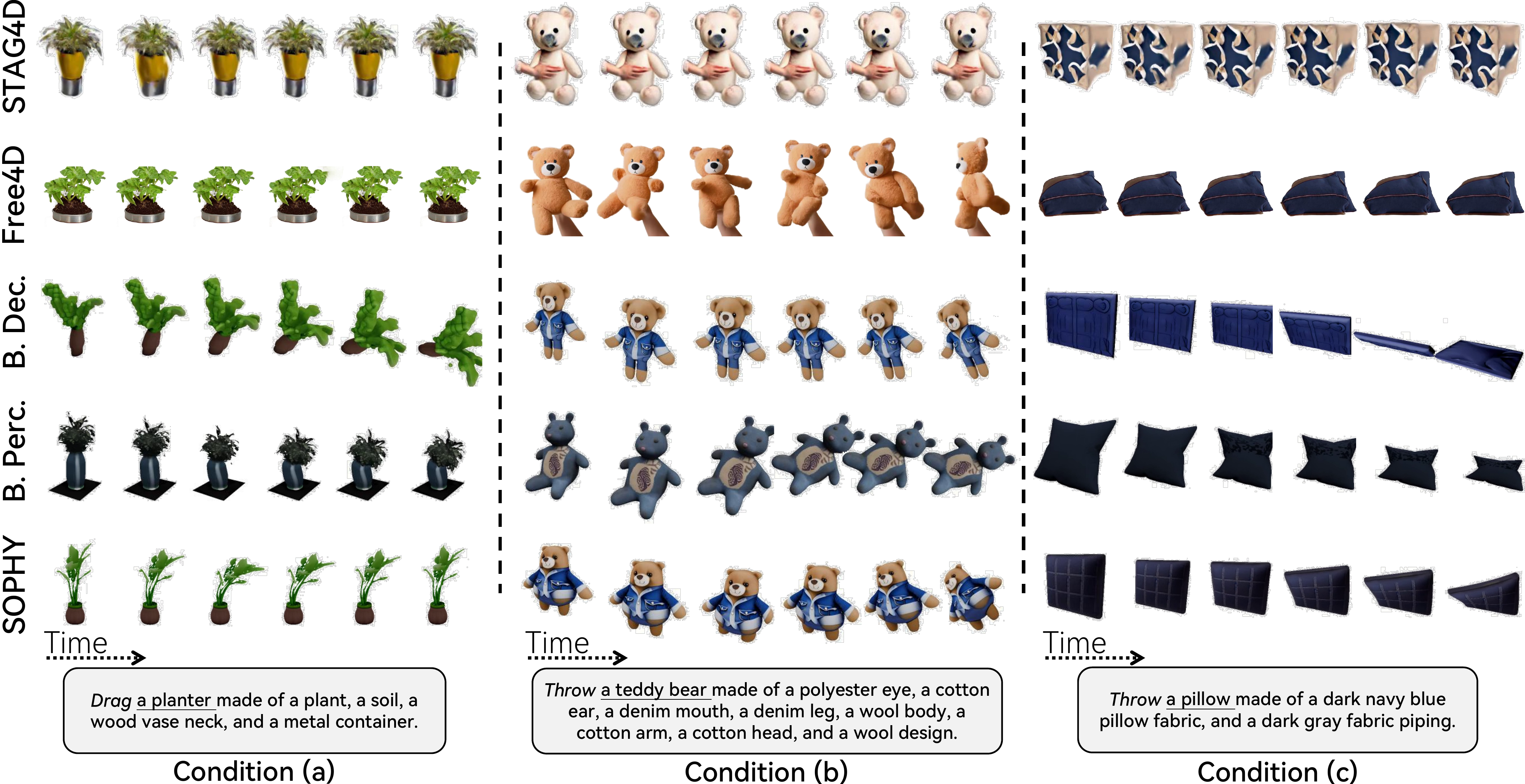}
    \caption{{\bf Qualitative comparison in the text-to-4D setting.}
    The text prompts used as input conditions are shown below. Note that the ground plane is not shown here as well.
    }    \label{fig:text_cond_results}
    \vspace{-6mm}
\end{figure*}

\subsection{Text-to-4D}
\label{sec:text-to-4D}

In this subsection, we evaluate\method~for text-to-4D generation. The evaluation is the same as before, with the only difference being the input condition (text instead of an image). We test on prompts that describe the object category, part tags, and fine-grained material categories for our test objects, following the prompts of Section \ref{sec:vlm_guidance}. We also include the test scenario in the prompt.

\paragraph{Competing methods \& Metrics.} We compare our method with STAG4D~\cite{zeng2024stag4d}, which generates deforming 3DGS driven by text prompts. We also consider Free4D~\cite{liu2025free4d} and the two baselines in the previous section here. The evaluation metrics are the same as in the image-to-4D setting.

\paragraph{Results.}
In~\Cref{tab:text_to_4d_metrics}, it is observed that\method~achieves the highest performance in VideoPhy and CLIP scores, demonstrating better alignment with real-world physics and given text prompts. 
We also present a visual comparison for the generated dynamics in~\Cref{fig:text_cond_results}. We observe that STAG4D fails to produce noticeable deformations, even for soft objects like teddy bears. Although Free4D can occasionally generate dynamic content as shown in~\Cref{fig:text_cond_results}(b), the content mismatches the conditional signal. For the two baselines, since they exclude material properties from the generation process, they fail to produce appropriate physical attributes aligned with the text prompts. For instance, in~\Cref{fig:text_cond_results}(a), ``B. Dec.'' generates a very \emph{stiff} plant while ``B. Perc.'' creates a \emph{soft} metal container.
Compared to them, our method produces more geometrically coherent and physically plausible results. 

\section{Conclusion}
\vspace{-1mm}
We presented a new generative model of 3D objects incorporating geometry, color, and physical material properties. Our experiments demonstrated significant benefits of the approach, including generating physically plausible 4D videos from images and text. Interesting avenues for future work include extending our generative model for synthesizing whole scenes and generating other phenomena \eg, fluids~\cite{li2024neural,muller2003particle,ummenhofer2019lagrangian} or gases~\cite{feldman2005animating}. More accurate simulators, \eg, based on finite elements~\cite{sifakis2012fem,vega,trusty2022mixed}, could be used instead since our model generates the full objects' volume. Our annotated dataset is only a first step towards more physics-aware 3D datasets -- enlarging it with more objects and material properties would be a fruitful direction.

\paragraph{Acknowledgments} This work has received funding from the European Research Council (ERC) under the Horizon research and innovation programme (No. 101124742).

\bibliographystyle{plain}
\bibliography{main}


\appendix

\newpage
\section{Additional Analysis}
\label{sec:supp_analysis}

\subsection{Conditioning on images from other datasets}
\begin{figure}[t]
    \centering
    \includegraphics[width=\linewidth]{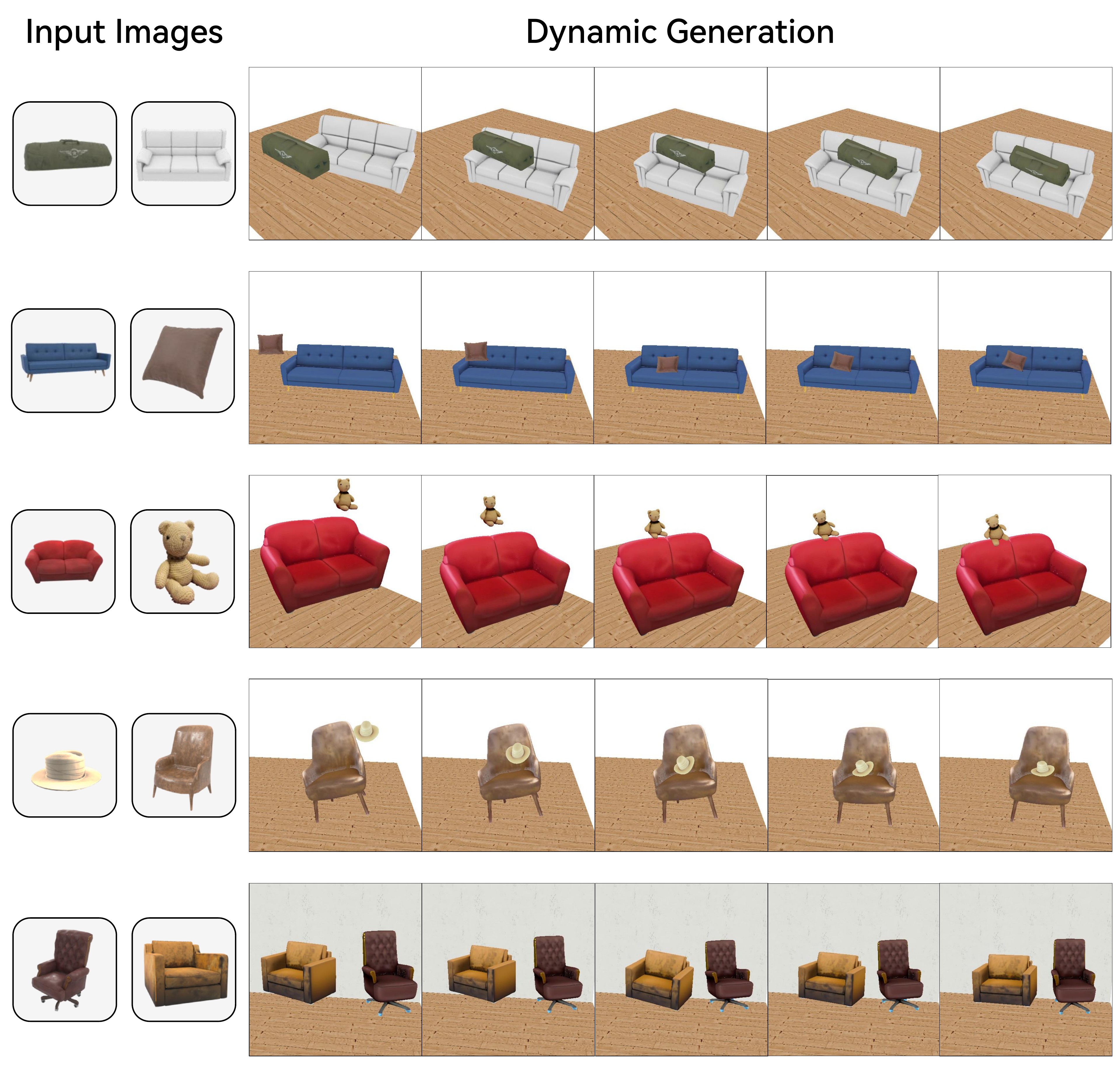}
    \caption{\textbf{Generated results by conditioning our method on images from Objaverse~\cite{deitke2023objaverse} shown on the left.}}
    \label{fig:generalization}
\end{figure}

\begin{figure}[h]
    \centering
    \includegraphics[width=\linewidth]{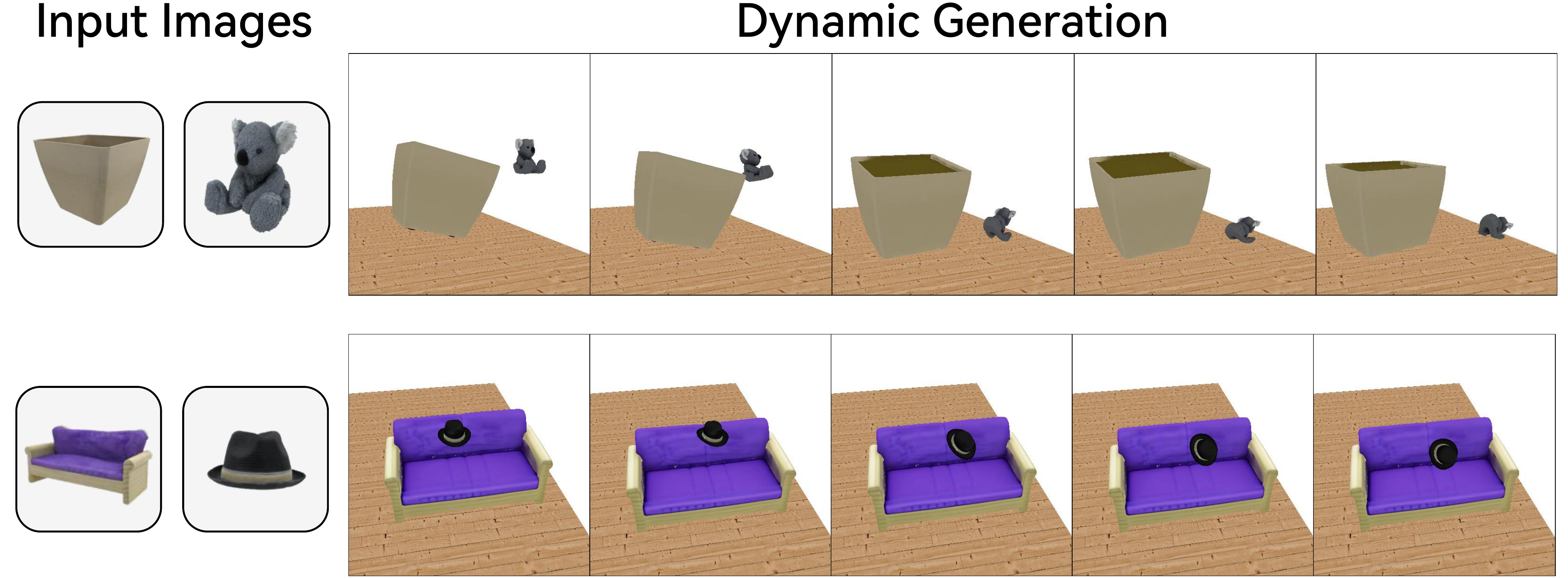}
    \caption{\textbf{Generated results by conditioning our method on images from Google Scanned Objects~\cite{downs2022google} shown on the left.}}
    \label{fig:generalization_gso}
\end{figure}

Here, we investigate how well\method~generalizes to conditioned images sourced from different datasets. 
We examine both synthetic and real-world scanned objects. Specifically, we choose the Objaverse~\cite{deitke2023objaverse}, a widely used synthetic dataset in previous 3D generation research, along with the Google Scanned Objects~\cite{downs2022google}, which is a high-quality dataset featuring scanned everyday items. We obtain 2D renderings of several 3D objects from them to serve as input conditional signals for our method. After completing the physics-aware 3D generation, we simulate the generated assets using the pipeline described in~\Cref{sec:verification} of the manuscript, finally obtaining dynamic videos. We present the results in~\Cref{fig:generalization} and~\Cref{fig:generalization_gso}. Even though our method was not specifically trained on these datasets, it can still generate 3D assets with plausible material properties and physical simulations. 

\begin{figure}[t]
    \centering
    \includegraphics[width=\linewidth]{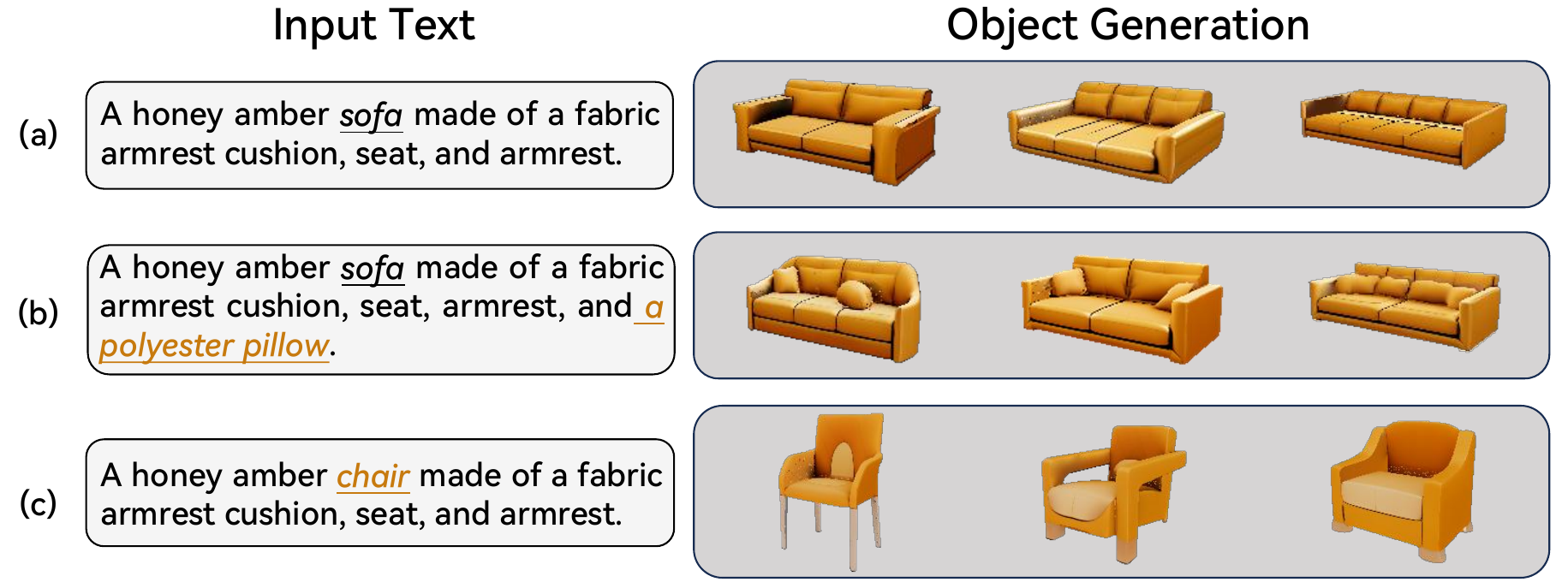}
    \caption{\textbf{Changing parts or object categories in text prompts shown on the left. On the right, we show different samples of objects generated by our method based on the edited prompts.}}
    \label{fig:text_control_1}
\end{figure}

\subsection{Conditioning on text prompt variations}

In this subsection, we study whether\method~can generate different 3D objects aligned well with the prompts based on edited text conditions. Specifically, we qualitatively demonstrate changes in the generated results by changing object parts or categories in the input prompt. The results are shown in~\Cref{fig:text_control_1}.
Starting with the text shown in~\Cref{fig:text_control_1}(a), we add a description of ``a polyester pillow'' in (b). The resulting generations accurately reflect this change in the descriptions. Furthermore, we alter the object category from ``sofa'' in (a) to ``chair'' in (c). As shown in~\Cref{fig:text_control_1}(c),\method~can generate similar shapes for the unedited parts in the prompt, such as ``cushion'' and ``armrest'', while the rest of the shape changes to reflect the edited category.

\section{Implementation Details}
\label{sec:supp_implementation_details}

\begin{figure}
    \centering
    \includegraphics[width=0.7\linewidth]{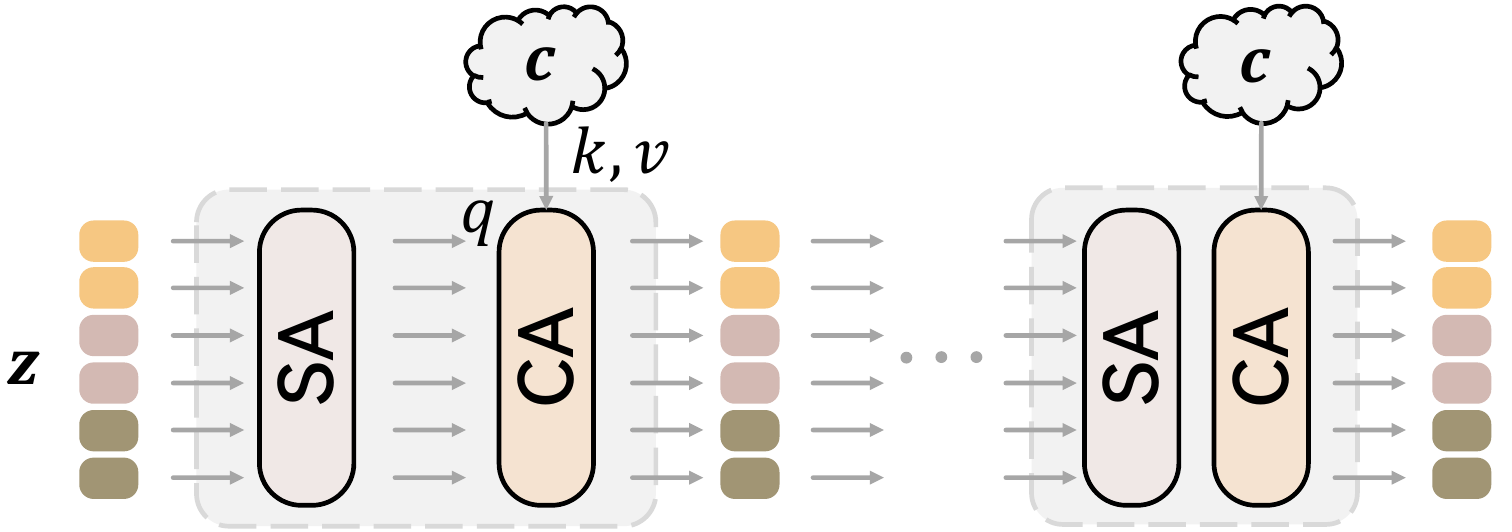}
    \caption{{\bf Denoiser architecture.} We adopt self attention (SA) and cross attention (CA) blocks to learn denoised representations starting from latents (see~\Cref{eq:reparameterization}). The cross attention incorporates conditions $\bc$ (images/prompts).}
    \label{fig:supp_denoiser}
\end{figure}

\subsection{Denoiser Architecture}
\label{sec:supp_denoiser}
In~\Cref{fig:supp_denoiser}, we present the architecture of the denoising network used in our method. We alternate self-attention (SA)  and cross-attention (CA) layers. The cross attention is used to inject the condition signal $\bc$ (image or text prompt) into the latent representation.

\subsection{Conditioning Signals for training}
\label{sec:supp_conditional_signals}
In the image-conditioned experiment, we use $10$ rendered images from randomly sampled viewpoints with a size of $256\times 256$ for training. In the text-conditioned experiment, each object is accompanied by $5$ text descriptions. $4$ of these descriptions are automatically generated using metadata from the 3DCoMPaT200~\cite{ahmed20253dcompat200}, as shown in the code snippet below. The fifth description is created by querying ChatGPT-4o~\cite{hurst2024gpt4o} for a text description of the object based on two rendered views.

\begin{lstlisting}[style=pythonstyle]
# Conditional Signals for Text
# 1. with color, coarse material type, and part label
text_1 = f"A {shape_name} made of"
for i, (part, mat, mat_fine) in enumerate(part_descriptions, start=1):
    mat_color = MAT_COLORS[mat].lower()
    if i != len(part_descriptions):
        text_1 += f" a {mat_color} {mat} {part},"
    else:
        text_1 += f" and a {mat_color} {mat} {part}."

# 2. with fine material type and part label
text_2 = f"A {shape_name} made of"
for i, (part, mat, mat_fine) in enumerate(part_descriptions, start=1):
    if mat_fine is None:
        mat_name = mat
    else:
        mat_name = mat_fine
    if i != len(part_descriptions):
        text_2 += f" a {mat_name} {part},"
    else:
        text_2 += f" and a {mat_name} {part}."

# 3. with part label
text_3 = f"A {shape_name} composed of"
for i, (part, mat, mat_fine) in enumerate(part_descriptions, start=1):
    if i != len(part_descriptions):
        text_3 += f" a {part_name},"
    else:
        text_3 += f" and a {part_name}."

# 4. with category label
text_4 = f"A 3D shape of {shape_name}."
\end{lstlisting}

\subsection{Data Preprocessing}
\label{sec:supp_data_preprocessing}
For object autoencoding and generation, we mainly follow the preprocessing pipeline of 3DShape2VecSet~\cite{zhang20233dshape2vecset} to obtain point-based shape representations. First, each 3D shape is converted into a watertight mesh using ManifoldPlus~\cite{huang2020manifoldplus} and is normalized to its bounding box. Then we sample a dense surface point cloud of 150K points. To train the network, we randomly sample 300K points in 3D space—each annotated with occupancy, color, and material properties—along with an additional set of 300K points from the near-surface region with the same attributes.
We apply the frequency embedding~\cite{mildenhall2020nerf,zhang20233dshape2vecset} to each point's position and color before forwarding it to the encoder. 
We note that color for off-surface points is assigned using a $1$-NN approach, where each point inherits the color of its nearest surface point. 

Material properties are determined through a label propagation process. 3DCoMPaT200~\cite{ahmed20253dcompat200} provides surface points with part labels for each object. To extend these labels to interior shape points, we query their $5$ nearest points and assign the label via a majority vote. Once part labels are obtained, we map corresponding material attributes based on part annotations in our material-augmented dataset.

\subsection{Perception-based Baseline}
\label{sec:supp_perception_based_baseline}
In~\Cref{sec:image-to-4D}, we introduced the perception-based baseline (\ie, ``B. Perc.''). The baseline uses perceptual models to estimate material properties based on an off-the-shelf 3D generation model. Here, we present more details of how this baseline is implemented. Specifically, we adopt TRELLIS~\cite{xiang2024structured}, a state-of-the-art 3D generation model, to generate textured 3D shapes given image or text conditions. To obtain the material properties of each generated shape, we then leverage an open-vocabulary 3D part segmentation model, Find3D~\cite{ma2024find}, to get the part labels for the sampled surface points. 
Note that the query part names we provide to Find3D are derived from the set of part labels in our dataset, which comes from 3DCoMPaT200~\cite{ahmed20253dcompat200}. Finally, we leverage ChatGPT-4o~\cite{hurst2024gpt4o} by providing it with two renderings of the generated 3D object and asking it to estimate the material properties for each part of the object retrieved by Find3D. The prompts we employed follow the one used in our dataset collection pipeline, as described in~\Cref{sec:supp_material_property_prompt}. After we acquire the material information for each part of the 3D object, we propagate this information to the points sampled within the object’s volume using the same strategy outlined in~\Cref{sec:supp_data_preprocessing}. Finally, we simulate the generated object with the estimated material properties.

\subsection{Training Details}
\label{sec:supp_training_details}

\begin{wrapfigure}[15]{r}{0.55\textwidth}
\centering
\vspace{-4mm}
\begin{minipage}{0.54\textwidth}
\centering
\includegraphics[width=\linewidth]{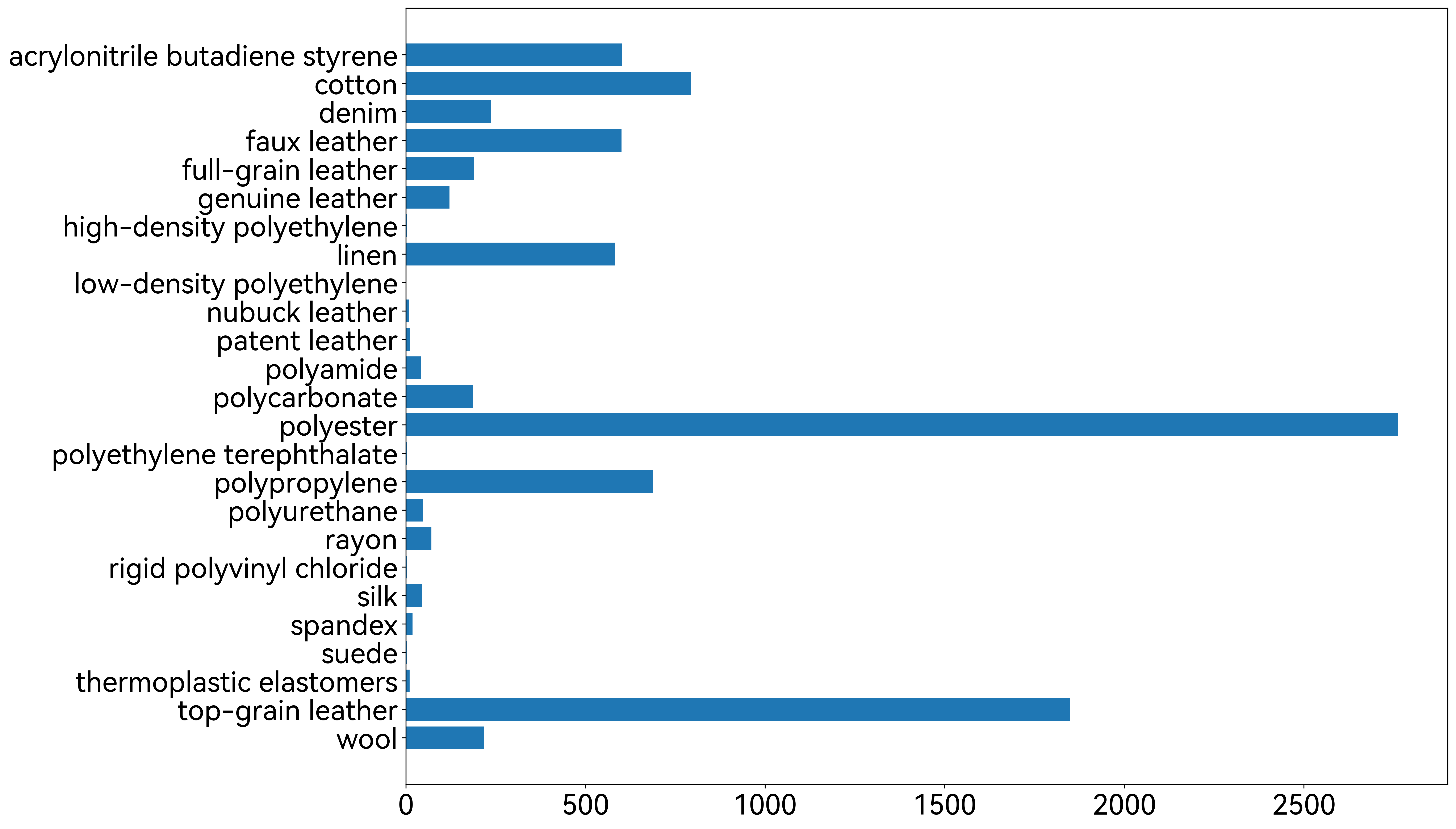}
\vspace{-2mm}
\caption{\bf Statistics of fine-grained material types in our dataset.}
\label{fig:sub_type_freq}
\end{minipage}
\end{wrapfigure}

For object autoencoding, we use a surface point cloud of $2,048$ points as input to the autoencoder. At each iteration, we sample $1,024$ query points from the bounding sphere of the object and another $1,024$ points from the surface region for attribute prediction. Following~\cite{zhang2025lagem}, we ensure an equal distribution of query points between occupied and non-occupied regions. 
We set the weight coefficients described in~\Cref{eq:loss} as $\lambda_c=\lambda_\nu=\lambda_\phi=\lambda_\rho=\lambda_M=0.1$, $\lambda_E=\lambda_\sigma=0.01$, and $\lambda_r=0.001$.
The autoencoder is trained with a batch size of $256$ for $500$ epochs, using a learning rate that linearly increases to $10^{-4}$
over the first $25$ epochs before gradually decaying to $10^{-5}$ following a cosine schedule. 

For object generation, we train our diffusion models with a batch size of $256$ for $3,000$ epochs. The learning rate is linearly increased to $2 \times 10^{-4}$ over the first $200$ epochs and then gradually decays to $10^{-6}$ using a cosine schedule. We adopt the default hyperparameter settings of EDM~\cite{karras2022elucidating}. During sampling, the final latent codes are obtained using only 18 denoising steps. All our training experiments are conducted on two NVIDIA L40s GPUs.

\section{Dataset Details}
\label{sec:supp_dataset_details}

\begin{table*}[t]
    \centering
    \small
    \setlength{\tabcolsep}{1mm}{
    \begin{tabular}{l|cccccccccccc|c}
    \toprule
        Split & Bag  &  Bed & Chair & Crib & Hat & Headband & Love Seat & Pillow & Planter & Sofa & Teddy Bear & Vase & Total\\
    \midrule
       Train &75&418&898&27&18&27&139&71&383&278&37&91& 2462 \\
        Valid &10&26&52&5&2&5&21&11&21&18&6&3&180 \\
       Test & 24 & 48 & 106 &8& 7&10 &39&  18 & 46 &  31 &11& 14& 362\\
   \bottomrule
    \end{tabular}
    }
    \caption{\bf Data composition of our proposed dataset.}
    \label{tab:dataset_composition}
\end{table*}

\begin{figure}[t]
    \centering
    \includegraphics[width=0.96\linewidth]{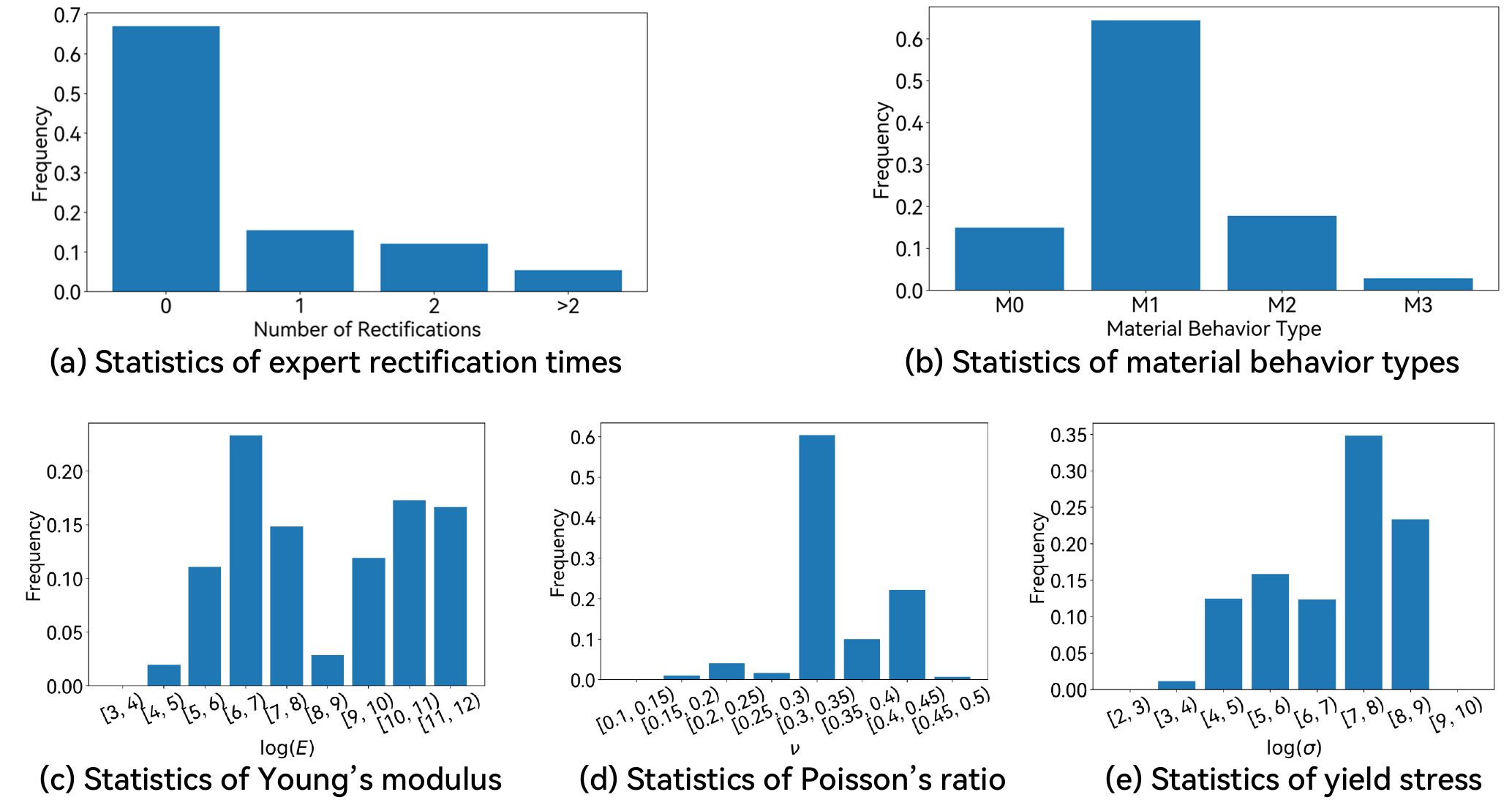}
    \caption{\textbf{More dataset statistics.}}
    \label{fig:dataset_stats}
\end{figure}

In~\Cref{tab:dataset_composition}, we present the detailed composition of the proposed dataset, which covers $12$ common categories with part-level material annotations for $3,004$ shapes. After the VLM-guided material annotation, we obtained $8$, $7$, $10$ types of fine-grained materials for ``fabric'', ``leather'', and ``plastic'' respectively\footnote{We did not obtain parts labeled with ``bonded leather'' for the ``leather'' type, and parts labeled with ``flexible polyvinyl chloride'', ``polystyrene'' for the ``plastic'' type from the VLM.}. The occurrence of parts labeled with these fine-grained materials is shown in~\Cref{fig:sub_type_freq}. 

\begin{figure}
    \centering
    \includegraphics[width=0.85\linewidth]{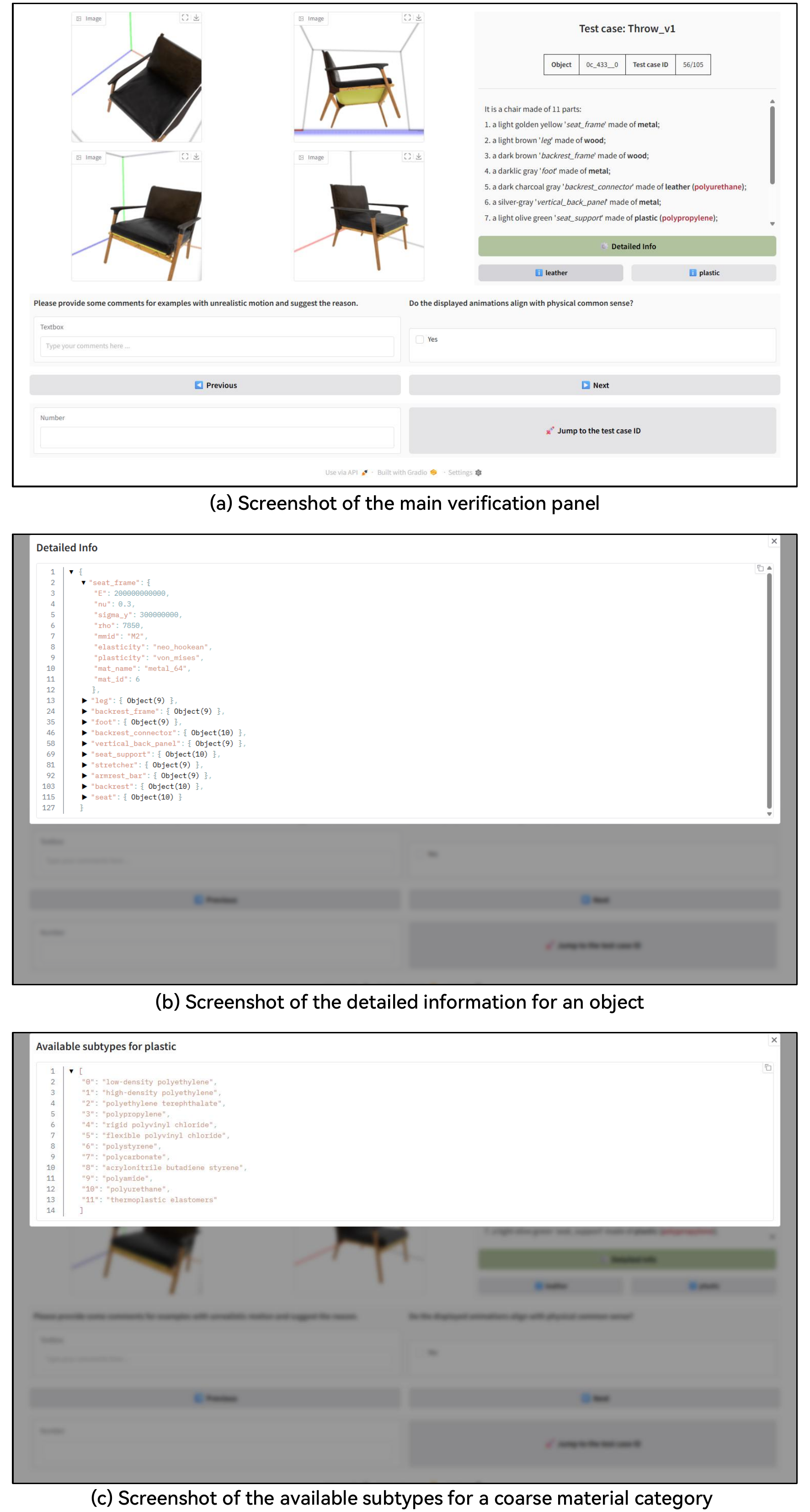}
    \caption{\textbf{Screenshots of the interactive panel for verifying material annotations.}}
    \label{fig:dataset_verification_panel}
\end{figure}

Furthermore, we present more dataset statistics in~\Cref{fig:dataset_stats}. We begin by illustrating the number of rectification attempts made by experts for each object during the data verification phase outlined in~\Cref{sec:verification}. As presented in~\Cref{fig:dataset_stats}(a), approximately one-third of the data has received at least one round of material parameter updates based on the experts' feedback\footnote{Please refer to~\Cref{fig:dataset_verification_panel}, where we provide some screenshots of the web interface we used for verifying material annotations.}. This highlights the importance of human verification in correcting the initial proposals generated by the VLM. Furthermore, it is noteworthy that most verifications are completed within two iterations, which demonstrates the efficiency of our data annotation pipeline. Next, we provide the statistics of several material properties, including material behavior types in~\Cref{fig:dataset_stats}(b), the logarithm of Young's modulus in~\Cref{fig:dataset_stats}(c), Poisson's ratio in~\Cref{fig:dataset_stats}(d), and the logarithm of yield stress in~\Cref{fig:dataset_stats}(e). Specifically, M0 represents pure elastic materials using neo-Hookean elasticity and identity plasticity. M1 represents plastic materials with softening implemented with neo-Hookean elasticity and von Mises plasticity with damage. M2 is for plastic material without softening, which is supported by neo-Hookean elasticity and von Mises plasticity. Finally, M3 stands for granular material with StVK elasticity and Drucker-Prager plasticity as the material models. These figures showcase the material diversity of our collected dataset.

In summary, our dataset offers detailed material properties, which serve as a first step for advancements in simulation-driven learning, physics-informed generative modeling, and interactive virtual environments. We hope that our dataset can expand the scope of 3D perception research and contribute to developing effective algorithms and techniques that improve our understanding of the physical world.

\begin{figure*}[t]
    \centering
    \includegraphics[width=1.0\linewidth]{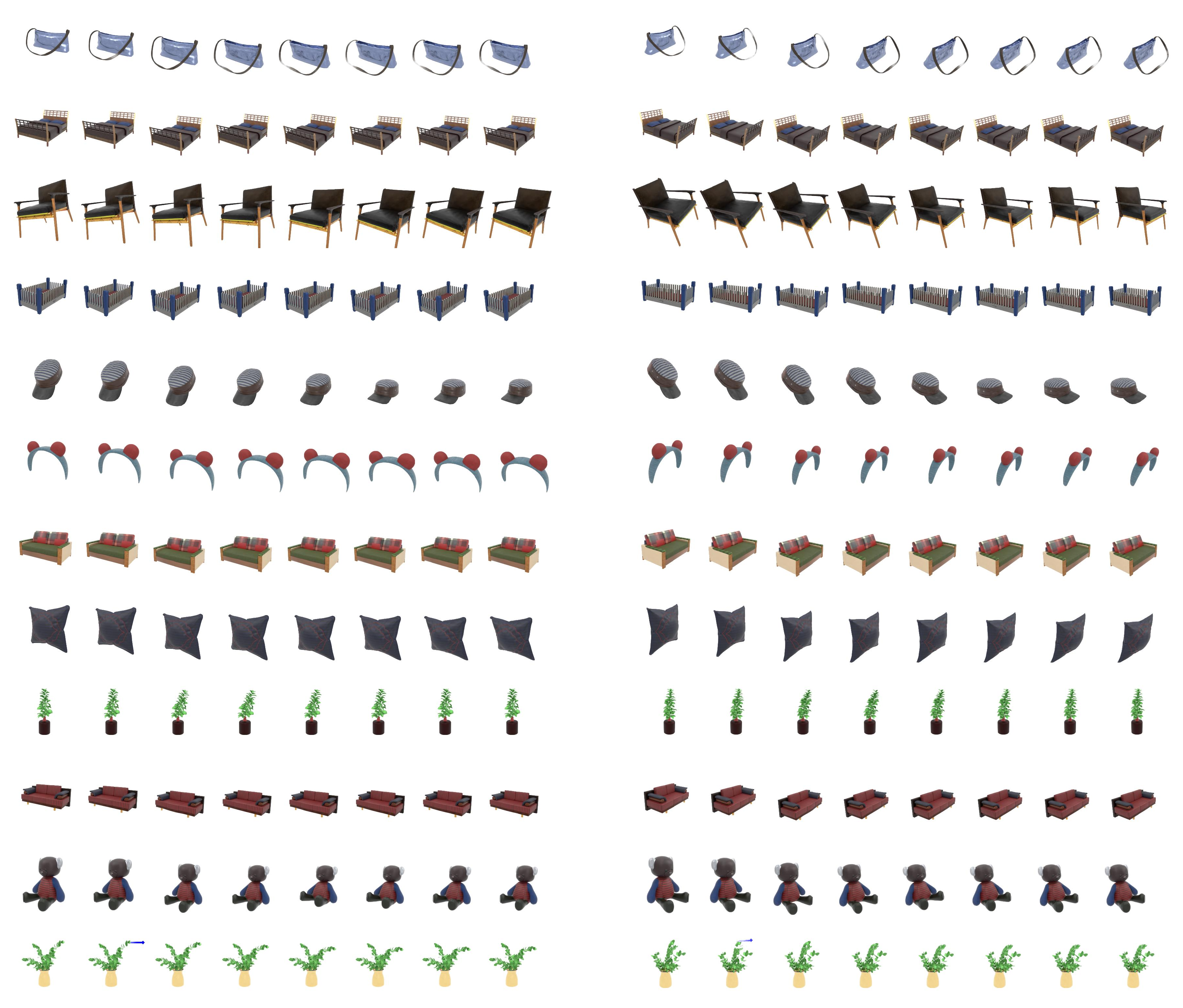}
    \caption{\bf Sampled simulation results from our proposed dataset.}
    \label{fig:dataset_showcase}
\end{figure*}

\section{Full Prompts}
\label{sec:supp_full_prompts}
In this section, we provide the detailed prompts we used in the VLM-guided material annotation stage described in~\Cref{sec:material_annotation}.

\subsection{Fine-grained Material Types}
\label{sec:supp_fine_grained_types}
We pre-define a list of available fine-grained material types for coarse materials, \ie, ``fabric'', ``leather'', and ``plastic'', from 3DCoMPaT200~\cite{ahmed20253dcompat200}. This is necessary due to the significant variations in material characteristics within these broader classifications. Concretely, the available fine-grained materials for ``fabric'' include: cotton, wool, polyester, silk, denim, spandex, linen, and rayon. 
The available fine-grained materials for ``leather'' include: full-grain leather, top-grain leather, genuine leather, nubuck leather, suede, patent leather, bonded leather, and faux leather.
The available fine-grained materials for ``plastic'' include: low-density polyethylene, high-density polyethylene, polyethylene terephthalate, polypropylene, rigid polyvinyl chloride, flexible polyvinyl chloride, polystyrene, polycarbonate, acrylonitrile butadiene styrene, polyamide, polyurethane, and thermoplastic elastomers.

When processing a part component that belongs to any of these coarse material types, we supply the VLM with the corresponding fine-grained material types to achieve a more precise result regarding its material composition, as shown below.

\begin{lstlisting}[style=markdownstyle]
You are an intelligent AI assistant for computer graphics, physical simulation, and material science. 

Follow the user's requirements carefully and make sure you understand them. 

Keep your answers short and to the point. 

Do not provide any information that is not required. 

You are going to identify the most likely fine-grained material type for one or more parts of the object in the attached image(s).

The attached images describe a [SHAPE NAME] made of [N_P] parts: [PART-MATERIAL DESCRIPTION].

Given the appearance and your knowledge on material composition, please choose the most suitable fine-grained material type for the part(s): [A LIST OF PART NAMES].

The available options for the [COARSE-GRAINED MATERIAL NAME] material type are: [A LIST OF AVAILABLE FINE-GRAINED MATERIAL NAMES].

Please provide your answer in the following JSON format:
```
{
  "part_name": "most_suitable_material_type",
  ...: ...  # other parts
}
'''
The output should **only** contain the dictionary.
\end{lstlisting}

\subsection{Material Models and Parameters}
\label{sec:supp_material_property_prompt}
\begin{lstlisting}[style=markdownstyle]
You are an intelligent AI assistant for computer graphics, physical simulation, and material science. 

Follow the user's requirements carefully and make sure you understand them. 

Keep your answers short and to the point. 

Do not provide any information that is not required. 

You are going to use the Material Point Method to simulate the motion of the object shown in the attached image(s). 

To simulate the effect, you need to specify an elastic and a plastic material model, along with material parameters such as Young's modulus, Poisson's ratio, and yield stress. 

Note that the material parameters should be reasonable for the object shown in the image(s). 

More specifically, when the material parameters are used to simulate dropping, throwing, or tilting the object, the object should behave according to physical common sense.


The available material models are list below. 
# Available elastic material models (with parameters required) 
1. Neo-Hookean elasticity (Young's modulus, Poisson's ratio); 
2. StVK elasticity (Young's modulus, Poisson's ratio). 
# Available plastic material models (with parameters required) 
1. Identity plasticity; 
2. von Mises plasticity (Young's modulus, Poisson's ratio, yield stress); 
3. Drucker-Prager plasticity (Young's modulus, Poisson's ratio, friction angle); 

The available combinations of material models for each material category are listed below. The leading number is the **Combination ID**.
# ceramic
M1. Neo-Hookean elasticity, von Mises plasticity with damage;
# fabric
M0. Neo-Hookean elasticity, Identity plasticity;
M1. Neo-Hookean elasticity, von Mises plasticity with damage;
# leather
M0. Neo-Hookean elasticity, Identity plasticity;
M1. Neo-Hookean elasticity, von Mises plasticity with damage;
# metal
M2. Neo-Hookean elasticity, von Mises plasticity;
# plant
M0. Neo-Hookean elasticity, Identity plasticity;
# plastic
M1. Neo-Hookean elasticity, von Mises plasticity with damage;
# soil
M3. StVK elasticity, Drucker-Prager plasticity;
# wood
M1. Neo-Hookean elasticity, von Mises plasticity with damage;

The attached image(s) describe the object you are going to simulate.

It is a [SHAPE NAME] made of [N_P] parts: [PART-MATERIAL DESCRIPTION].

For each part, you need to specify **both the elastic and the plastic** material model and the material parameters, such as Young's modulus, Poisson's ratio, density, etc. 

Please provide your answer in the following JSON format (For Young's modulus and yield stress, the unit is Pa. For density, the unit is kg/m^3.): 
```
{
"part_name": {
    "CID": "Mx",  // Combination ID
    "E": youngs_modulus,
    "nu": poissons_ratio,
    "...": ...,  // other parameters, e.g., yield stress ("sigma_y"), friction angle ("phi"), density ("rho")
}
..., // other parts
}
'''
The output should **only** contain the dictionary. 
\end{lstlisting}

\subsection{Update with Expert Feedback}
 
In this case, we send three messages with the role set by \texttt{user}, \texttt{assistant}, \texttt{user} in succession. The first \texttt{user} prompt is the same as the prompt for obtaining the initial material properties. The \texttt{assistant} prompt is the output of the original query from the VLM. The second \texttt{user} prompt is:

\begin{lstlisting}[style=markdownstyle]
The original output creates unrealistic dynamics when the object [TEST CASE DESCRIPTION] in the simulator.

Specifically, [USER COMMENT].

Given this information, please update the material parameters to make the object behave more realistically.

The output should be formatted as the original version.
\end{lstlisting}
\section{Limitations}
\label{sec:supp_limitations}

Our current dataset of annotated shapes includes 3K objects and 15.5K labeled parts. This size is relatively small, especially for training large-scale 3D or 4D foundation models. We believe that our semi-automatic pipeline can be enhanced in the future to generate significantly larger datasets. 
Artifacts in 3D shape and texture generation are sometimes visible in our method (\eg, see the left, dark-shaded armrest of the first sofa shown in Figure \ref{fig:text_control_1}). 
Generating alternative representations based on our physics-aware latents and denoiser, such as meshes \cite{xiang2024structured} or marching tetrahedra \cite{shen2021deep}, along with the generation of high-resolution textures and physically based rendering materials (PBR) \cite{siddiqui2024meta}, in an end-to-end manner could help reduce artifacts and further enhance the quality of the results.

\end{document}